\documentclass[a4paper,11pt]{article}
\usepackage{amssymb}
\usepackage{abstract}
\usepackage{tabulary}
\usepackage{geometry}
\usepackage{fancyhdr}
\usepackage{color}
\usepackage{indentfirst}
\usepackage{bm}
\usepackage{amssymb,amsmath}
\usepackage{makecell,rotating,multirow,diagbox}
\usepackage{colortbl}
\usepackage{url}
\usepackage{authblk}
\usepackage{booktabs}
\usepackage{tabularx}

\usepackage{array}
\usepackage{makecell}
\usepackage{lipsum}
\usepackage{etoolbox}
\usepackage{natbib}
\usepackage{subfig}

\usepackage{algorithm}
\usepackage{algorithmic}
\usepackage{enumerate}
\usepackage{graphicx}
\usepackage{xcolor}
\usepackage{multirow}
\usepackage{textcomp,booktabs}
\usepackage{setspace}
\usepackage{lineno}
\usepackage{hyperref}
\usepackage{authblk}

\title{Efficient Bayesian estimation for GARCH-type models via Sequential Monte Carlo}
\author[a,b]{Dan Li \thanks{ Corresponding author. \protect\\ {\it{E-mail address:}} d31.li@hdr.qut.edu.au (Dan Li).}}
\author[c]{Adam Clements}
\author[a,b]{Christopher Drovandi}

\affil[a]{School of Mathematical Sciences, Queensland University of Technology, Australia}
\affil[b]{Australian Research Council Centre of Excellence for Mathematical \& Statistical Frontiers (ACEMS)}
\affil[c]{School of Economics and Finance,Queensland University of Technology, Australia}

\date{}

\begin{document}

\maketitle

\begin{abstract}

The advantages of sequential Monte Carlo (SMC) are exploited to develop parameter estimation and model selection methods for GARCH (Generalized AutoRegressive Conditional Heteroskedasticity) style models. It provides an alternative method for quantifying estimation uncertainty relative to classical inference. Even with long time series, it is demonstrated that the posterior distribution of model parameters are non-normal, highlighting the need for a Bayesian approach and an efficient posterior sampling method. Efficient approaches for both constructing the sequence of distributions in SMC, and leave-one-out cross-validation, for long time series data are also proposed. Finally, an unbiased estimator of the likelihood is developed for the Bad Environment-Good Environment model, a complex GARCH-type model, which permits exact Bayesian inference not previously available in the literature.
\end{abstract}


\setlength{\abovedisplayskip}{4.5pt}
\setlength{\belowdisplayskip}{4.5pt}

\newpage
\section{Introduction}

In many financial applications, volatility, defined as the conditional standard deviation of asset returns, is an important quantity. Many financial applications are based on volatility forecasts, such as risk management, asset pricing and portfolio optimization \citep{engle2001garch}. Therefore modelling of volatility has attracted a lot of research attention. One of the most influential models that describes the dynamics of volatility is the AutoRegressive Conditional Heteroscedasticity (ARCH) model by \citet{engle1982autoregressive} and the Generalized ARCH (GARCH) model by \citet{bollerslev1986generalized}, which is a more parsimonious extension of ARCH. The standard GARCH model of \citet{bollerslev1986generalized} has successfully captured many characteristics of financial asset returns, and forms the basis for a wide range of expressions used to capture various empirical features of returns data. For instance, in order to capture the stylised fact of the asymmetric effect \citep{black1976studies}, a range of models such as the exponential GARCH (EGARCH) model \citep{nelson1991conditional}, the Glosten-Jagannathan-Runkle GARCH model (GJR-GARCH) of \citet{glosten1993relation}, the absolute value GARCH (AGARCH) model of \citet{hentschel1995all}, and the Bad Environment-Good Environment (BEGE) model \citep{bekaert2015bad} have been developed. Given the wide range of GARCH type models, it is crucial to develop an efficient statistical inference framework to select between these models.

Frequentist inference based on maximum likelihood estimation (MLE) is commonly employed with the GARCH class of models. A limitation of MLE is that it only produces point estimates of parameters, with the uncertainty of these parameter estimates based on asymptotic theory where parameter estimates are assumed to be normally distributed \citep{hamilton1994time}. With the development of new financial instruments based on volatility, such as volatility options, modelling the full predictive density of volatility is attracting attention (e.g., \citealp{corradi2009predictive, corradi2011predictive}; and \citealp{griffin2016bayesian}). To produce such distributions, here we develop a Bayesian approach to statistical inference for the GARCH class of models. Under this approach, the unknown parameters are regarded as random variables, whose posterior distributions may be far from normal. In contrast to the confidence interval under classical statistics, the credible interval of Bayesian statistics is easily interpreted and gives a more direct and intuitive probability statement about uncertainty in the parameter estimates, \citep{o2004bayesian}. Furthermore, our Bayesian approach can also provide full predictive distributions of the conditional volatility required for pricing volatility instruments.

An issue with Bayesian estimation is developing an efficient method for sampling from the posterior distribution based on Monte Carlo methods \citep{robert2004monte}. The most commonly used approach is the Markov Chain Monte Carlo (MCMC) method \citep{metropolis1953equation,hastings1970monte}, which has been utilised for Bayesian analysis of the GARCH class of models (see, for example, \citealp{kim1998stochastic}; \citealp{vrontos2000full}; and \citealp{takaishi2009adaptive}). Despite the popularity of MCMC, it can be costly and difficult to tune the associated parameters when using MCMC \citep{roberts2009examples} with conventional MCMC algorithms becoming less effective when the target distribution is far from normally distributed. In more recent times, the sequential Monte Carlo (SMC) method for static Bayesian models \citep{chopin2002sequential,del2006sequential} has been devised as a complementary method to MCMC. SMC traverses a set of particles through a sequence of probability distributions, which starts with one that is easy to sample from and ends with the ultimate target posterior distribution. The sequence of probability distributions can be attained either using a data annealing strategy \citep{chopin2002sequential} or the method of likelihood annealing \citep{neal2001annealed}. SMC iteratively applies three main steps to traverse the population of particles, which include re-weighting, re-sampling, and moving (or mutation). SMC is parallelisable and easy to adapt to become more efficient compared with the MCMC method. It is possible to use parallelisation to save computation costs especially when the likelihood function is computationally intensive, which we demonstrate is easily achievable for the GARCH class of models. In addition, SMC is efficient at exploring irregular posterior distributions, often found with GARCH type models. The intermediate samples created by SMC embedded with data annealing strategy can be reused to perform posterior predictions. Also, the estimate of evidence, which is a by-product from SMC methods, can be used in Bayesian model selection.

In this paper, we utilise SMC methods to develop a novel fully Bayesian approach for the widely used GARCH class of models. This Bayesian framework can be used for estimation, prediction and model selection, and generating a full predictive density for conditional volatility. Our Bayesian SMC approach provides an attractive alternative to the MCMC or frequentist approaches. This paper also contributes to the SMC literature as we demonstrate that SMC data annealing is useful for leave-one-out cross-validation when dealing with time series data. Finally, we develop an unbiased likelihood estimator for the BEGE model, which permits exact Bayesian inference for this model. \citet{bekaert2015bad} use an approximation to the likelihood, which we show produces biased posterior distributions. \citet{south2019sequential} use the BEGE model with biased likelihood as an example to illustrate a new SMC approach for parameter estimation.  Our paper is a comprehensive demonstration of the advantages that SMC can produce for GARCH-type models.

In Section 2, we provide an introduction to GARCH models and present the three models that are investigated in the paper. The methods for deriving an unbiased estimator for the likelihood of the BEGE model are also described. Section 3 provides a brief explanation of Bayesian statistics and the general SMC framework is given in detail. An efficient strategy to choose the sequence of intermediate probability distributions is also discussed. Section 4 provides a comparison of one-step forward predictions of volatility that are derived from the fully Bayesian approach and the classical approach, respectively. A leave-one-out cross-validation method for time series data is also described for evaluating the predictive performance of the forecasted volatility. In Section 5 a Bayesian model selection tool that is used to choose between candidate models is provided. Finally, a summary of the results is provided in Section 6 and concluding discussion is given in Section 7.

\section{GARCH models}

The standard GARCH model introduced by \citet{bollerslev1986generalized} successfully established a dynamic model of time-varying volatility which captures some of the important characteristics of financial asset returns \citep{engle1982autoregressive}. However, the classical GARCH model does not account for the stylised feature of the asymmetric effect \citep{black1976studies}. The GJR-GARCH \citep{glosten1993relation} model is an improvement over the classical GARCH model, and \citet{engle1993measuring} suggest that the model captures the asymmetric effect in stock returns best among other popular nonlinear models such as the EGARCH \citep{nelson1991conditional} model. Originally, the assumption of Gaussian innovations in the standard GARCH and GJR-GARCH model fails to generate a non-Gaussian distribution for returns, which is one of the most important properties of financial time series \citep{francq2011garch}. To solve this problem to some extent, the Student-t distribution suggested by \citet{bollerslev1987conditionally} is employed here. Besides, we also consider an alternative model that accommodates conditional non-Gaussianity and was based on the GJR-GARCH model, which is called bad environment-good environment (BEGE) model \citep{bekaert2015bad}.

\subsection{Classical GARCH(1,1) model}
GARCH models are observation driven, where the conditional variance is a function of not only its past values, but also past innovations to returns. To begin, define the innovations in returns, $u_{t}$ as a shock to returns:
\begin{align}
&r_{t}=\mu+u_{t},\rm \ and\\
&(u_t|\sigma_t) \sim f_\nu(u_t|\sigma_t) = \frac{\Gamma\left[\frac{1}{2}(\nu+1)\right]}{\pi^{\frac{1}{2}}\Gamma\left[\frac{1}{2}\nu\right]}\left[\left(\nu-2\right)\sigma_t^2\right]^{-\frac{1}{2}}\left[1+\frac{u_t^2}{\left(\nu-2\right)\sigma_t^2}\right]^{-\frac{1}{2}\left(\nu+1\right)},
\end{align}
where $r_{t}$ is the time series of asset returns with fixed mean $\mu$, with $u_{t}$ having a conditional Student-t distribution and a time varying conditional variance $\sigma_t^2$, with different GARCH models defining different processes for $\sigma_t^2$. In what follows, we present three models within the GARCH family considered in this paper.

Under the classical $\rm{GARCH}$$(p,q)$ model, the conditional variance is given by:
\begin{equation}
\sigma_{t}^{2}=\alpha_0+\sum_{j=1}^q\alpha_j u_{t-j}^2+\sum_{i=1}^p\beta_i\sigma_{t-i}^2,\qquad t\in\mathbb{Z},
\end{equation}
where $\alpha_0>0$, $\alpha_j>0$, $\beta_i>0$, $p>0$, $q>0$, $i=1,\dots,p$, $j=1,\dots,q$, and the stationary condition requires that $\sum_{j=1}^q\alpha_j+\sum_{i=1}^p\beta_i<1$ \citep{bollerslev1986generalized}. The wide success of the parsimonious GARCH$(1,1)$ model in extensive empirical applications makes this model the workhorse of financial applications (see, for example, \citealp{engle2001garch}; \citealp{alexander2006normal}; and \citealp{francq2011garch}). The GARCH$(1,1)$ model considered here is given by:
\begin{align}
\sigma_{t}^{2}=\alpha_0 +\beta_1\sigma_{t-1}^2+\alpha_1u_{t-1}^2,
\end{align}
where $\alpha_0$, $\alpha_1$, and $\beta_1$ are subject to the constraints discussed above in the general $\rm{GARCH}$$(p,q)$ model. The log-likelihood function of the GARCH$(1,1)$ model is given by:
\begin{equation}
\begin{aligned}
L(\bm{r}|\bm{\theta}_\mathrm{Garch})=&\sum_{t=1}^T \log{\left[\Gamma\left(\frac{\nu+1}{2}\right)\right]} - \log{\left[\Gamma\left(\frac{\nu}{2}\right)\right]} - \frac{1}{2}\left[\log{\left(\pi\left(\nu-2\right)\sigma_t^2\right)}\right]\\&- \frac{1}{2}\left[\left(\nu+1\right)\log{\left(1+\frac{u_t^2}{\left(\nu-2\right)\sigma_t^2}\right)}\right],
\end{aligned}
\end{equation}
where $\bm{r}$ is the observed time series of asset returns, and $\bm{\theta}_\mathrm{Garch}$ is a parameter set of the GARCH$(1,1)$ model.

\subsection{GJR-GARCH(1,1) model}
The asymmetric GJR-GARCH$(1,1)$ model is a GARCH-type model with a threshold indicator function, and has a different expression on conditional volatility $\sigma_t^2$ from the classical GARCH model, which is defined as:
\begin{align}
&\sigma_t^2=\alpha_0+\beta_{\sigma}\sigma_{t-1}^2+\phi u_{t-1}^2+\phi^{-}u_{t-1}^2(I_{u_{t-1}<0}),
\end{align}
where $I$ is a dummy variable which takes the value of $1$ if the innovation is negative, and $0$ otherwise; and $\alpha_0>0$, $\beta_{\sigma}>0$, $\phi^{-}>0$ and $\phi>0$ to ensure the conditional variance is positive and $\beta_{\sigma}+\phi+0.5\phi^{-} < 1$ to ensure the conditional variance is stationary. In this model, the negative shocks increase the conditional variance relative to positive shocks due to the asymmetric effect in stock returns. The log-likelihood function of the GJR-GARCH$(1,1)$ model has the same expression with the one of the GARCH$(1,1)$ model provided in Equation (5).

\subsection{Bad Environment-Good Environment (BEGE) model}
The BEGE model is based on non-Gaussian innovations and is defined as:
\begin{equation}
\begin{aligned}
&u_{t}=\sigma_{p}\omega_{p,t}-\sigma_{n}\omega_{n,t}\rm ,\ where\\
&\omega_{p,t} \sim \widetilde{\Gamma}(p_{t} , 1)\rm,\ and\\
&\omega_{n,t} \sim \widetilde{\Gamma}(n_{t} , 1)\rm,
\end{aligned}
\end{equation}
where $\widetilde{\Gamma}(k, \theta)$\footnote{The probability density function of $\widetilde{\Gamma}(k, \theta)$ is $f(x)=\frac{1}{\Gamma(k)\theta^k}(x+k\theta)^{k-1}\exp\left(-\frac{1}{\theta}(x+k\theta)\right)$ for $x>-k\theta$.} is the so-called centered gamma distributions \citep{bekaert2015bad} with a zero mean. The innovation $u_t$ is a linear combination of a bad environment component $\omega_{n,t}$ with a shape parameter of $n_t$, and a good environment component $\omega_{p,t}$ with a shape parameter of $p_t$. The conditional shock distribution of $u_t$ is generated from two gamma-distributed shocks. The time-varying shape parameters $n_t$ and $p_t$ are given by:
\begin{align}
&p_t=p_0+\rho_pp_{t-1}+\frac{\phi_p^{+}}{2\sigma_p^2}u_{t-1}^2I_{u_{t-1}\ge0}+\frac{\phi_p^{-}}{2\sigma_p^2}u_{t-1}^2(1-I_{u_{t-1}\ge0}), \rm \ and\\
&n_t=n_0+\rho_nn_{t-1}+\frac{\phi_n^{+}}{2\sigma_n^2}u_{t-1}^2I_{u_{t-1}\ge0}+\frac{\phi_n^{-}}{2\sigma_n^2}u_{t-1}^2(1-I_{u_{t-1}\ge0}).
\end{align}

The overall conditional variance $\sigma_t^2$ of $u_{t}$ in the BEGE model is:
\begin{align}
\sigma_t^2 \equiv var_t(r_{t})=\sigma_p^2p_t+\sigma_n^2n_t,
\end{align}
which is derived from the moment generating function of the centered gamma distribution \citep{bekaert2015bad}. As with the conditional variance, moments of higher order such as conditional skewness and excess kurtosis can be easily derived with tractable expressions. The conditional skewness is given by $\mathrm{skw_t}(r_t) = 2\left(\sigma_p^3p_t-\sigma_n^3n_t\right)/\sigma_t^{3}$, and the conditional kurtosis is given by $\mathrm{kur_t}(r_t) = 6\left(\sigma_p^4p_t+\sigma_n^4n_t\right)/\sigma_t^{4}$.
These higher-order conditional moments are time-varying compared with traditional asymmetric volatility GARCH models. The log-likelihood function of the BEGE model is written as:
\begin{align}
L(\bm{r}|\bm{\theta}_\mathrm{BEGE})=\sum_{t=1}^T\log{f_{\mathrm{BEGE}}(r_t|r_{t-1},\dots,r_1;\bm{\theta}_\mathrm{BEGE})},
\end{align}
where $\bm{\theta}_\mathrm{BEGE}$ is a parameter set of the BEGE model, and the conditional likelihood of observation $r_t$ $f_{\mathrm{BEGE}}(r_t|r_{t-1},\dots,r_1;\bm{\theta}_\mathrm{BEGE})$ is approximated numerically according to \citet{bekaert2015bad}. The cumulative distribution function (CDF) of $f_{\mathrm{BEGE}}(r_t|r_{t-1},\dots,r_1;\bm{\theta}_\mathrm{BEGE})$ at two points just above and below the observation $r_t$ are numerically evaluated. Then a finite difference approximation method is used to find the derivative of the CDF, which approximates the probability density function of $r_t$. As a result of the nature of approximation in this deterministic numerical method, the approximated likelihood of $r_t$ is biased. Below, we propose an unbiased estimation for the likelihood of $r_t$ of the BEGE model with importance sampling.

\subsubsection{Unbiased estimation for the BEGE likelihood}
According to \cite{bekaert2015bad}, the BEGE-distributed variable $u_t$ can be rewritten as
$u_{t}= \omega_{p,t}-\omega_{n,t}$, where $\omega_{p,t} \sim \widetilde{\Gamma}(p_{t} , \sigma_{p})$, $\omega_{n,t} \sim \widetilde{\Gamma}(n_{t} , \sigma_{n})$, and $u_t = r_t - \mu$.
Then the BEGE density can take the form as
\begin{equation}
\begin{aligned}
f_{\mathrm{BEGE}}(u_t) &= \int_{\omega_{p,t}}f_{\bm{u_t}}(u_t|\omega_{p,t})f(\omega_{p,t})\mathrm{d}\omega_{p,t}\ \\
&=\int_{\omega_{p,t}}f_{\omega_{n,t}}(\omega_{p,t} - u_t)f(\omega_{p,t})\mathrm{d}\omega_{p,t}.
\end{aligned}
\end{equation}
An unbiased estimator of $f_{\mathrm{BEGE}}(u_t)$ can be constructed using Monte Carlo methods. Firstly, we sample $M$ independent samples $\omega_{p,t}^1,\dots,\omega_{p,t}^M$ from $f(\omega_{p,t})$, then $f_{\mathrm{BEGE}}(u_t)$ can be estimated as
\begin{equation}
\begin{aligned}
\hat{f}_{\mathrm{BEGE}}(u_t) = \frac{1}{M} \sum_{i=1}^M f_{\omega_{n,t}}(\omega_{p,t}^i - u_t), \mathrm{where}\ \omega_{p,t}^i \stackrel{\text{i.i.d.}}{\sim} f(\omega_{p,t}), i=1,\dots,M.
\end{aligned}
\end{equation}
One main issue of the Monte Carlo integration method is that a large number of Monte Carlo draws, $M$, may be required to increase the accuracy of the Monte Carlo estimates, which makes it more computationally intensive. Therefore, we adopt importance sampling to reduce variance of Monte Carlo estimates and improve computational efficiency. When it is hard to sample from our target distribution $h(\cdot)$ directly, we can construct an importance distribution, $g(\cdot)$, which is easy to sample from. The following identity is the basis for importance sampling
\begin{equation}
\begin{aligned}
\mathop{\mathbb{E}_{g}}\left[\frac{h(x)}{g(x)}\right] &= \int \frac{h(x)}{g(x)} g(x) \mathrm{d}x = \int h(x) \mathrm{d}x \\
&\approx \frac{1}{M} \sum_{i=1}^M \frac{h(x_i)}{g(x_i)}, x_i \stackrel{\text{i.i.d.}}{\sim} g(\cdot).
\end{aligned}
\end{equation}
In the case of the BEGE model, the target distribution is $h(\omega_{p,t}) = f_{\omega_{n,t}}(\omega_{p,t} - u_t)f(\omega_{p,t})$, and the importance sampling estimator of $f_{\mathrm{BEGE}}(u_t)$ is given by
\begin{equation}
\begin{aligned}
\hat{f}_\mathrm{IS}(u_t) = \frac{1}{M} \sum_{i=1}^M \frac{h(\omega_{p,t}^i)}{g(\omega_{p,t}^i)}, \omega_{p,t}^i \stackrel{\text{i.i.d.}}{\sim} g(\omega_{p,t}),
\end{aligned}
\end{equation}
where a sound proposal distribution $g(\omega_{p,t}^i)$ needs to be selected. The process of constructing an efficient importance distribution for estimating $f_{\mathrm{BEGE}}(u_t)$ is provided in the Appendix. The general importance sampling estimator in (14) is unbiased.

\citet{andrieu2009pseudo} show that using an unbiased likelihood estimator within MCMC produces a Markov chain that converges to the idealised posterior that assumes the likelihood can be computed exactly. \citet{chopin2013smc2} and \citet{duan2015density} show that using an unbiased likelihood estimator in an SMC algorithm for data annealing and likelihood annealing, respectively, produces an algorithm that also targets the exact posterior. We adopt a Bayesian statistical framework for parameter estimation, model selection and prediction for GARCH-type models that is described in the next section.

\section{Bayesian Inference}

\subsection{Introduction to Bayesian Statistics}
Bayes' theorem captures learning from observed data and experience. Here, $\bm{y}$ represents the observed data and $\bm{\theta}$ represents the set of parameters of a model believed to have generated $\bm{y}$.
The posterior distribution of $\bm{\theta}$ is given by:
\[\pi(\bm{\theta}|\bm{y})=\frac{f(\bm{y}|\bm{\theta})\pi(\bm{\theta})}{Z},\]
where ${f(\bm{y}|\bm{\theta})}$ is the likelihood function, ${\pi(\bm{\theta})}$ is the prior, and ${Z}=f(\bm{y})=\int_{\bm{\theta}} f(\bm{y}|\bm{\theta})\pi(\bm{\theta})\mathrm{d}\bm{\theta}$ is the normalising constant of the posterior that is independent of $\bm{\theta}$.

When our interest is in generating samples of $\bm{\theta}$ from the posterior, the normalising constant ${Z}$ can be ignored and then the posterior function of $\bm{\theta}$ can be expressed as $\pi(\bm{\theta}|\bm{y})\propto f(\bm{y}|\bm{\theta})\pi(\bm{\theta})$.
Given that directly sampling from the posterior distribution $ \pi(\bm{\theta}|\bm{y})$ is generally infeasible, the Markov chain Monte Carlo (MCMC)  method \citep{metropolis1953equation,hastings1970monte} is the most widely used sampling method.  One standard MCMC algorithm  is the Metropolis-Hastings (MH) algorithm, which is straightforward to implement.
Under the MH algorithm, a proposal distribution $q(\bm{\theta^*}|\bm{\theta})$ shall be carefully chosen, where the new sample candidate value $\bm{\theta^*}$ that depends on the current state $\bm{\theta}$ is proposed with probability
\[\alpha(\bm{\theta} \to \bm{\theta{^*}})=\min\left(1,\frac{f(\bm{y}|\bm{\theta^*})\pi(\bm{\theta^*})q(\bm{\theta}|\bm{\theta^*})}{f(\bm{y}|\bm{\theta})\pi(\bm{\theta})q(\bm{\theta^*}|\bm{\theta})}\right).\]
The reader is referred to \citet{chib1995understanding} for the theoretical details of MH algorithms. According to \cite{andrieu2009pseudo}, the exact likelihood $f(\bm{y}|\bm{\theta})$ can be replaced by its unbiased estimator $\hat{f}(\bm{y}|\bm{\theta})$, with the corresponding MCMC algorithm still converging to the posterior distribution $\pi(\bm{\theta}|\bm{y})$.

However, the popular MCMC algorithm is not free from limitations. For the MH algorithm, an optimal proposal distribution $q(\bm{\theta^*}|\bm{\theta})$ can lead to better Markov chains that converge more quickly and efficiently explore the parameter space. But it may be costly to tune the proposal distribution. In order to improve the efficiency of MCMC algorithms, adaptive MCMC methods have been studied in the literature (e.g., \citealp{haario2001adaptive,haario2006dram}; \citealp{roberts2009examples}). However, the Markov property of the process may be destroyed with many adaptive strategies and optimal acceptance rates targeted by some adaptive methods are based on strong posterior distribution assumptions, which are unlikely to hold in many applications. Moreover, when the target distribution is multimodal, the conventional MCMC algorithm may become stuck in a local mode for many iterations. An alternative method named sequential Monte Carlo (SMC) was designed, which could be viewed as a complementary method to MCMC. SMC has several advantages over MCMC and a more detailed description of this method is given in the next section.

\subsection{Sequential Monte Carlo}

\subsubsection{Sequential Monte Carlo framework}
\noindent Here, we are focusing on the use of SMC for static Bayesian models \citep{chopin2002sequential}. In SMC, a set of $N$ weighted particles, $\left\{W_t^i, \bm{\theta}_t^i\right\}_{i=1}^N$, are traversed through a sequence of posterior distributions. Therefore, we need to build a sequence of distributions $\pi_t(\bm{\theta}|\bm{y})$ for $t=0,\dots,T$, which explore the ultimate target distribution gradually and where $\pi_T(\bm{\theta}|\bm{y})$ equals $\pi(\bm{\theta}|\bm{y})$. One method to construct the sequence of distributions is the data annealing strategy \citep{chopin2002sequential}. Under this method the target sequence is created as $\pi_t(\bm{\theta}|\bm{y}_{1:t})\propto f(\bm{y}_{1:t}|\bm{\theta})\pi(\bm{\theta})$, where $\bm{y}_{1:t}$ represents the observation data up to time $t$. Another method called likelihood annealing \citep{neal2001annealed} can also be used to generate the sequence when the whole observed dataset is already available. The sequence formed by the likelihood annealing method is given by:
\[ \pi_t (\bm{\theta}|\bm{y})=\frac{f(\bm{y}|\bm{\theta})^{\gamma_t}\pi(\bm{\theta})}{Z_t} \propto f(\bm{y}|\bm{\theta})^{\gamma_t}\pi(\bm{\theta}),\]
where $\gamma_t$ is referred to as the temperature and $0=\gamma_0\le \dots \le \gamma_t\le \dots \le \gamma_T=1$. We use both of the methods in this study, however, the data annealing approach is preferred when data are collected sequentially. A more detailed analysis and comparison between these two sequence constructions is demonstrated in Section 3.2.2. As with MCMC, unbiased likelihood estimators in SMC and exact Bayesian inference (up to Monte Carlo error) can be produced \citep{chopin2013smc2,duan2015density}.

In order to generate a properly weighted sample for target $t$, we need to reweight the particle set at target $t-1$, $\left\{W_{t-1}^i, \bm{\theta}_{t-1}^i\right\}_{i=1}^N$. The reweighting step can be achieved by importance sampling using:
\[w_t^i=W_{t-1}^i\tilde{w}_t^i=W_{t-1}^i\frac{\eta_t(\bm{\theta}_{t-1}^i|\bm{y})}{\eta_{t-1}(\bm{\theta}_{t-1}^i|\bm{y})},\]
\citep{chopin2002sequential}, for $i=1,\dots,N$ where $w_t^i$ needs to be normalised, and $\eta_t(\cdot)$ is the unnormalised version of $\pi_t(\cdot)$. The so-called incremental weight $\tilde{w}_t^i$ is calculated by dividing the unnormalised target density at $t$, $\eta_t(\bm{\theta}_{t-1}^i|\bm{y})$, by the unnormalised target density at $t-1$, $\eta_{t-1}(\bm{\theta}_{t-1}^i|\bm{y})$. Unfortunately, the reweighting step generally leads to reductions in the weighted samples efficiency, which can be measured by the effective sample size (ESS; \citealp{liu2008monte}). ESS refers to the number of perfect samples among all the $N$ generated samples and is computed by:
\[\mathrm{ESS}_t=\frac{N}{1+\mathrm {CV}[w_t(\bm{\theta})]},\]
where CV[$w_t(\bm{\theta})$] is the coefficient of variation of the unnormalised weights of target $t$. The ESS can be approximated using the normalized weights as $1/\sum_{i=1}^N(W_t^i)^2$. Therefore, when the ESS becomes too small, we resample the particle values proportional to their weights. This resets the approximate ESS back to $N$.

However, the resampling strategy will result in duplication, especially for particle values with relatively large weight. A moving step is required to increase the number of unique particles. The moving step is usually the most computationally expensive part of the SMC process, and the MCMC kernels of invariant distribution $\pi_t$ are commonly employed in the literature. Generally, in static Bayesian models $R_t$ iterations of a multivariate normal random walk MCMC kernel \citep{chopin2002sequential} are taken, the approach adopted here is designed to diversify the particles. We propose to use the MH MCMC algorithm. It is important to note that the population of particles can be used to automatically tune the covariance matrix of the multivariate normal random walk. As the proposals of particles can be rejected by the MCMC kernel, it is then necessary to repeat the MCMC kernel $R_t$ times to maintain a target level of diversity.

According to \cite{salomone2018unbiased}, $R_t$ can be determined adaptively by choosing an optimal tuning scale for the covariance matrix of the MCMC kernel. The optimal scale is selected based on expected square jumping distance (ESJD; \citealp{sherlock2009optimal}). We randomly assign each particle one of several tuning scales and we estimate the ESJD for each particle using a single MCMC iteration. The tuning scale which resulted in the highest median ESJD value is selected as optimal. The ESJD is given by:
\[\text{ESJD}=\alpha(\bm{\theta}_t^i \to \bm{\theta}{^*})\Lambda_t^i=\alpha(\bm{\theta}_t^i \to \bm{\theta}{^*})(\bm{\theta}_t^i - \bm{\theta}{^*})^T\hat{\Sigma}_t^{-1}(\bm{\theta}_t^i - \bm{\theta}{^*}),\]
which is the product of the MH acceptance ratio and the Mahalanobis square jumping distance \citep{salomone2018unbiased}, $i=1,\dots,N$, and $t=0,\dots,T$. After obtaining the optimal scale, we keep applying an MCMC kernel to all the particles until a target number of particles' ESJD values are higher than a threshold. We set this threshold at the median value of the Mahalanobis distances between each pair of particles before the move step. As can be seen, the total number of MCMC iterations can be adaptively attained after meeting the threshold. The SMC sampling method is summarized in Algorithm 3.1.

\begin{algorithm}
	\setcounter{algorithm}{0}
	\renewcommand{\thealgorithm}{\arabic{section}.\arabic{algorithm}}	
	\caption{SMC Sampler}	
	\begin{algorithmic}[1]
		\STATE Sample $\bm{\theta}_0^i \sim \pi(\cdot)$ and set $W_0^i=1/N$ for $i=1,\dots,N$
		\FOR{$t=1,\dots,T$}
		\STATE Reweight: $w_t^i=W_{t-1}^i\frac{\eta_t(\bm{\theta}_{t-1}^i|\bm{y})}{\eta_{t-1}(\bm{\theta}_{t-1}^i|\bm{y})}$
		\STATE Set $\bm{\theta}_t^i=\bm{\theta}_{t-1}^i$ for $i=1,\dots,N$
		\STATE Normalization of the weights: $W_t^i=w_t^i/\sum_{k=1}^{N}w_t^k$ for $i=1,\dots,N$
		\STATE Computation of ESS: $\text{ESS}=1/\sum_{i=1}^N(W_t^i)^2$
		\IF{$\text{ESS}<\alpha N$, with $\alpha \in (0,1]$}
		\STATE Resample the particles and set $W_t^i=1/N$, producing a new weighted particle set $\left\{W_t^i,\bm{\theta}_t^i\right\}_{i=1}^N$
		\STATE Move each particle with an MCMC kernel distribution of invariant distribution $\pi_t$ for $R_t$ iterations
		\ENDIF
		\ENDFOR

	\end{algorithmic}
\end{algorithm}

\subsubsection{Efficient sequence construction method}
\noindent Under the SMC approach, the sequence of target distributions can be defined through both likelihood annealing and data annealing. These two annealing methods provide similar estimates of posterior distributions for the parameters of the three GARCH-type models, which are shown in Section 6.1. Under the data annealing strategy, the sequence of target distributions are obtained smoothly when the observations arrive one at a time. The sequence of posterior distributions constructed by using data annealing method are given by: \[\pi_t(\bm{\theta}|\bm{y}_{1:t}) \propto f(\bm{y}_{1:t}|\bm{\theta})\pi(\bm{\theta}),\]
where $t=0,\dots,T$, and $\bm{y}_{1:t}$ represents the data available up to current time $t$. The corresponding unnormalised weight of the particle $i$ at time $t$ is thus:
\[w_t^i=W_{t-1}^i\frac{f(\bm{y}_{1:t}|\bm{\theta}_{t-1}^i)}{ f(\bm{y}_{1:t-1}|\bm{\theta}_{t-1}^i)}= W_{t-1}^i f(\bm{y}_t|\bm{y}_{1:t-1} ,\bm{\theta}_{t-1}^i),\]
which needs to be normalised. It is important to note that to update the weights only the conditional likelihood of the introduced data point needs to be computed. As more returns become available, the re-weighting step in data annealing does not need to re-process previously collected data. This provides extensive computational savings, especially when the time series is long.
It is worth noting that all data still need to be traversed from the beginning when moving the particles with an MCMC kernel. Therefore, it is not an online method but some benefits can be provided from the re-weighting step.

The data annealing strategy is also useful for performing posterior prediction. One way to consider if a GARCH-type model is appropriate would be to focus on the model's ability to forecast. Our Bayesian approach provides a clear method to deduce the posterior predictive distribution conditional on currently available data. The intermediate posterior distribution $\pi_{t-1}(\bm{\theta}|\bm{y}_{1:t-1})$ from data annealing is utilised when forecasting one-step ahead posterior prediction for data at time $t$. Meanwhile, the corresponding posterior predictive density that is used to measure the prediction accuracy can be estimated. Section 4 details the procedure of estimating the posterior predictive distribution.

\section{Bayesian posterior predictive distribution}
\noindent In this section, the fully Bayesian and traditional classical approaches are compared in the context of generating one-step ahead forecasts of conditional variances. The distribution of the one-step ahead conditional variances together with 95\% prediction intervals are offered by the fully Bayesian approach, which cannot be provided by the classical approach. We also provide a fully Bayesian approach for assessing time series models' predictive performance, which can be achieved easily with SMC data annealing.

\subsection{Bayesian posterior predictive distribution}
Under the fully Bayesian approach, the posterior predictive distribution for the unknown observations can be produced. New observations $\hat{y}_{t+1}$ can be drawn from the posterior predictive distribution given the original data $y_{1:t}$ up to date $t$:
\[\mathrm{P}(\hat{y}_{t+1}|y_{1:t})=\int_{\bm{\theta}}\mathrm{P}(\hat{y}_{t+1}|\bm{\theta},y_{1:t})\pi_t(\bm{\theta}|y_{1:t})\mathrm{d}\bm{\theta}.\]

The posterior predictive distribution accounts for the uncertainty associated with the parameters without making asymptotic assumptions. Here, we consider one-step ahead forecasts for the posterior predictions of the conditional variance $\hat{\sigma}^2_{t+1}$ given its past values $\sigma_{1:t}^2$ and the real data $y_{1:t}$ up to time $t$. 

The posterior predictive distribution is particularly easy to estimate via SMC with data annealing.  Denote the SMC weighted sample from the posterior $\pi_t(\bm{\theta}|y_{1:t})$ as $\left\{W_t^i, \bm{\theta}_t^i\right\}_{i=1}^N$. Then, the posterior predictive distribution can be approximated by $\left\{W_t^i,(\hat{\sigma}^2_{t+1})^i\right\}_{i=1}^N$ where $(\hat{\sigma}^2_{t+1})^i \sim p(\sigma^2_{t+1}|y_{1:t},\sigma_{1:t}^2,\bm{\theta}_t^i)$ for $i=1,\ldots,N$.  From the weighted sample, a point estimate (e.g.\ mean or median) and predictive interval can be easily estimated.  The computation associated with constructing the approximation to the predictive posterior can be easily vectorised and parallelised.

\subsection{Forecast through maximum likelihood estimation (MLE) approach }
We compare the Bayesian SMC approach for forecasting with MLE. In order to form one-step ahead forecasts with MLE, at time $t$ we first find the maximum likelihood estimate $\bm{\theta}_\mathrm{MLE}^t$ based on the observations up to time $t$. MLE parameter estimates are obtained by maximising each model's log-likelihood provided in Section 2 via numerical optimisation. After obtaining the MLE parameter estimates at time $t$, a point estimate of the prediction of the conditional variance from the three GARCH-type models at time $t+1$ can be obtained by utilizing one of Equations (4), (6) or (10) in Section 2.

\subsection{Forecast evaluation}
In order to assess which model provides more accurate forecasts, we adopt the leave-future-out cross-validation (LFO-CV) method \citep{burkner2019approximate}, which is a modification of the popular leave-one-out cross-validation (LOO-CV) method.
LOO-CV is a well-known approach to assess out-of-sample predictive performance using in-sample fits. For an out-of-sample data set $y_{t+1},\dots,y_{T}$ and a in-sample data set $y_{1},\dots,y_{t}$, LOO-CV requires to refit the model $T-t$ times based on all the data except the forecast data point on which the predictive performance is assessed.
\citet{burkner2019approximate} state that LOO-CV is not suitable for assessing predictive performance when the data is sequentially ordered in time since future observations may provide some information that affects in-sample model fit, and thus the predictions may be affected if only leaving one observation out at a time. Therefore, when making one-step ahead forecasts at time $t+1$ we use LFO-CV that leaves out all the future values after time $t$ to assess predictive performance for time series models.

To measure the predictive accuracy, the expected log predictive density (elpd; \citealp{vehtari2017practical}) is approximated by:
\[\mathrm{elpd}_{\mathrm{LFO}}= \sum_{t=\tau-1}^{T-1}\log{p(y_{t+1}|y_{1:t})} ,\]
\citep{burkner2019approximate} where $\tau$ is time point that we decide to start assessing the predictions, $T$ is the number of observations in the dataset, and the density $p(y_{t+1}|y_{1:t})$ can be approximated by evaluating an estimated density, which is constructed using kernel density estimation based on the weighted sample $\left\{W_t^i,\hat{y}_{t+1}^i\right\}_{i=1}^N$, at the true observation $y_{t+1}$. The standard approach to LFO-CV would be to re-compute the posterior distribution (referred to here as a refit) as each new data point is introduced, however, this would be computationally intensive. \citet{burkner2019approximate} propose an algorithm to approximate LFO-CV and reduce the number of refits reducing computation time. However, our data annealing SMC naturally re-estimates the posterior distribution in a computationally efficient manner as more data are introduced, and hence, we demonstrate it is valuable for performing LFO-CV.
The weighted sample from the posterior $\left\{W_t^i, \bm{\theta}_t^i\right\}_{i=1}^N$ based on different subsets of observations $y_{1:t}$ is a by-product of posterior simulation via data annealing SMC, which means there is no need to refit the time series model again to perform and test predictions.

A simulation study and an empirical application are performed in Section 6.3 to evaluate the predictive performance for conditional variance and real returns, respectively, and explore the relative merits of the fully Bayesian approach relative to the MLE approach. To assess predictive accuracy we provide an estimated $\mathrm{elpd}_{\mathrm{LFO}}$ 
for each candidate model with different datasets.
We also illustrate that the predictive distribution of the conditional volatility can provide us with different levels of confidence about the magnitude of the volatility at the next time point.

\section{Model choice}
Practitioners are not only interested in single model analysis, but also the relative performance of competing models. Under a fully Bayesian framework, the relative performance of one model against another candidate model is assessed using the Bayes factor \citep{jeffreys1935some,kass1995bayes}. The Bayes factor compares two different models, $m_1$ and $m_2$, with parameter vectors $\bm{\theta}_{m_1}$ and  $\bm{\theta}_{m_2}$ is given by:
\[\mathrm{BF}_{m_1,m_2}=\frac{P(\bm{y}|m_1)}{ P(\bm{y}|m_2)}=\frac{\int{P(\bm{y}|\bm{\theta}_{m_1},m_1)P(\bm{\theta}_{m_1}|m_1)} \mathrm{d}\bm{\theta}_{m_1}}{\int{P(\bm{y}|\bm{\theta}_{m_2},m_2)P(\bm{\theta}_{m_2}|m_2)} \mathrm{d}\bm{\theta}_{m_2}},\]
which is just the ratio of normalising constants under each model. The normalising constant is also referred to as the marginal likelihood or evidence. The larger the evidence in favour of model $m_1$ relative to $m_2$ is, the greater the Bayes factor $\mathrm{BF}_{m_1,m_2}$. One challenge of implementing Bayes factor is the intensive computation of the evidence. A review of some commonly used methods of estimating the model evidence is given in \citet{friel2012estimating}.

Fortunately, an estimate of the evidence can be obtained as a by-product from SMC methods. The estimate of evidence $Z$ can be computed as $Z=Z_T/Z_0=\prod_{t=1}^TZ_t/Z_{t-1}$ with $Z_0=1$ \citep{del2006sequential}. Under likelihood annealing SMC, the ratio of normalizing constants $ Z_t/Z_{t-1}$ is given by:
\[\frac{Z_t}{Z_{t-1}}=\int_{\bm{\theta}}f(\bm{y}|\bm{\theta})^{\gamma_t-\gamma_{t-1}}\pi_{t-1}(\bm{\theta}|\bm{y})\mathrm{d}\bm{\theta}.\]
Then the estimate of log evidence $\log{Z}$ is given by:
\begin{equation*}
\begin{aligned}
\log{Z}&=\sum^T_{t=1}\log{\mathbf{E}_{\pi_{t-1}}[ f(\bm{y}|\bm{\theta})^{\gamma_t-\gamma_{t-1}}]}\\&\approx\sum^T_{t=1}\log{\left(\sum_{i=1}^Nw_t^i\right)},
\end{aligned}
\end{equation*}	
where $w_t^i$ is the unnormalised weight from Algorithm 3.1. A similar estimator under data annealing can be easily derived.
It should be noted that due to the adaptive techniques employed in the SMC methods, \citet{beskos2016convergence} show that the estimator of log evidence is consistent but not unbiased.
The estimates of log evidence of the three GARCH models obtained through both likelihood annealing and data annealing SMC are provided in Section 6.4.

Other widely used model selection tools such as the Bayesian information criterion (BIC) \citep{schwarz1978estimating}, which provides a crude approximation to the logarithm of Bayes factor \citep{kass1995bayes}, and deviance information criterion (DIC) \citep{spiegelhalter2002bayesian} have restrictive assumptions and limitations. For instance, the BIC will behave poorly when the implicit prior assumed by BIC is not a multivariate normal distribution \citep{kuha2004aic}, and DIC is only valid when the posterior can be approximated well with a multivariate normal distribution \citep{gelman2014understanding}. However, the fully Bayesian approach based on the Bayes factor does not rely on the restrictive approximations. A simulation study provided in Section 6.4 demonstrates the Bayesian model evidence's ability to select the true model for the motivating GARCH-type models of interest in this paper.

\section{Results}
The plan of this section is as follows. Section 6.1 compares the results for the BEGE model with the biased and unbiased likelihood.
In Section 6.2, the performance of the MCMC and SMC methods are compared, and the results for the posterior distributions for the GARCH-type models are presented. The one-step ahead forecasts of conditional variances by using both classical MLE and Bayesian approaches are illustrated in Section 6.3. The model selection results obtained from the fully Bayesian approach are presented in Section 6.4. Computer code for implementing our methods is available at \url{https://github.com/DanLAct/GARCH-SMC}.

In this study we use $N=10,000$ particles in SMC to generate posterior distributions. The real time series data are monthly log stock returns for the S\&P 500 Composite Index from July 1926 to January 2018 from the Center of Research in Securities Prices (CRSP). The parameters in the GARCH-type models we aim to estimate respectively are $\bm{\theta}_{\mathrm{Garch}(1,1)}=(\mu^\mathrm{G}, \alpha_0^\mathrm{G}, \alpha_1, \beta_1, \nu^\mathrm{G})$, $\bm{\theta}_{\mathrm{GJR}}=(\mu^{\mathrm{GJR}}, \alpha_0^{\mathrm{GJR}}, \beta_{\sigma}, \phi, \phi^{-}, \nu^\mathrm{GJR})$, and $\bm{\theta}_{\mathrm{BEGE}}=(\mu^{\mathrm{B}}, p_0, n_0, \rho_p, \rho_n, \phi_p^+, \phi_n^+, \phi_p^-, \phi_n^-, \sigma_p, \sigma_n)$. Some constraints on the parameters are imposed initially, so that the priors are set as below.
The priors on the parameters of the $\rm{GARCH}(1,1)$ model are given by:
\begin{equation*}
\begin{aligned}
&&\alpha_0^\mathrm{G} \quad &\sim \quad \mathcal{U}(0, 0.3),&
&&\alpha_1 \quad &\sim\quad \mathcal{U}(0, 0.5),&
&&\beta_1 \quad &\sim \quad\mathcal{U}(0, 0.99),&\\
&&\mu^\mathrm{G} \quad&\sim\quad \mathcal{U}(-0.9, 0.9),& \rm \ and
&&\left(\nu^\mathrm{G}-2\right) \quad &\sim \quad \Gamma(2,3),&
\end{aligned}
\end{equation*}
where $\mathcal{U}(\mathord{\cdot})$ denotes a uniform probability density function, $\Gamma(a,b)$ represents a gamma distribution with shape $a$ and scale $b$, and the degrees of freedom $\nu^\mathrm{G} > 2$. To satisfy stationarity it is necessary to enforce $\alpha_1+\beta_1\le0.9999$.
The priors on the parameters of the GJR-GARCH$(1,1)$ model are given by:
\begin{equation*}
\begin{aligned}
&&\alpha_0^{\mathrm{GJR}}, \phi, \phi^- \quad &\sim \quad \mathcal{U}(0, 0.3),&
&&\beta_{\sigma} \quad &\sim\quad \mathcal{U}(0, 0.99),&\\
&&\mu^{\mathrm{GJR}} \quad&\sim\quad \mathcal{U}(-0.9, 0.9),&\rm \ and
&&\left(\nu^\mathrm{GJR}-2\right) \quad &\sim \quad \Gamma(2,3).&
\end{aligned}
\end{equation*}
To satisfy stationarity we need $\phi+\frac{1}{2}\phi^-+\beta_{\sigma}\le0.9999$.
The priors on the parameters of the BEGE model are given by:
\begin{equation*}
\begin{aligned}
&&p_0, \phi_p^+, \phi_p^- \quad &\sim \quad \mathcal{U}(0, 0.5),&\\
&&\sigma_p, \sigma_n \quad &\sim\quad \mathcal{U}(0, 0.3),&
&&\rho_p, \rho_n\quad &\sim \quad\mathcal{U}(0, 0.99),&\\
&&n_0 \quad&\sim\quad \mathcal{U}(0, 1),&
&&\phi_n^+ \quad&\sim\quad \mathcal{U}(-0.2, 0.1),&\\
&&\phi_n^- \quad&\sim\quad \mathcal{U}(0, 0.75),&\rm \ and
&&\mu^\mathrm{B} \quad&\sim\quad \mathcal{U}(-0.9, 0.9).&
\end{aligned}
\end{equation*}
Additionally, to satisfy stationarity we require that $\rho_p + \frac{1}{2}\phi_p^++\frac{1}{2}\phi_p^-\le0.995$ and $\rho_n + \frac{1}{2}\phi_n^++\frac{1}{2}\phi_n^-\le0.995$. All parameters are assumed to be independent in the prior.

\subsection{Comparison of Biased and Unbiased BEGE}
The simulated data generated from the BEGE model are used to study the comparative performance of the biased and unbiased likelihood of the BEGE model. Figure 1 demonstrates that the approximate likelihood of $f_{\mathrm{BEGE}}(u_t)$ can lead to biased posteriors, especially for $\phi_p^-$ and $\mu^B$. Even though the unbiased estimators of the likelihood produced by Monte Carlo integration and importance sampling produce similar posterior distributions, the importance sampling estimating method is more efficient.
\begin{figure}[!htp]
	\centering
	\includegraphics[width=1.05\linewidth, height=10cm]{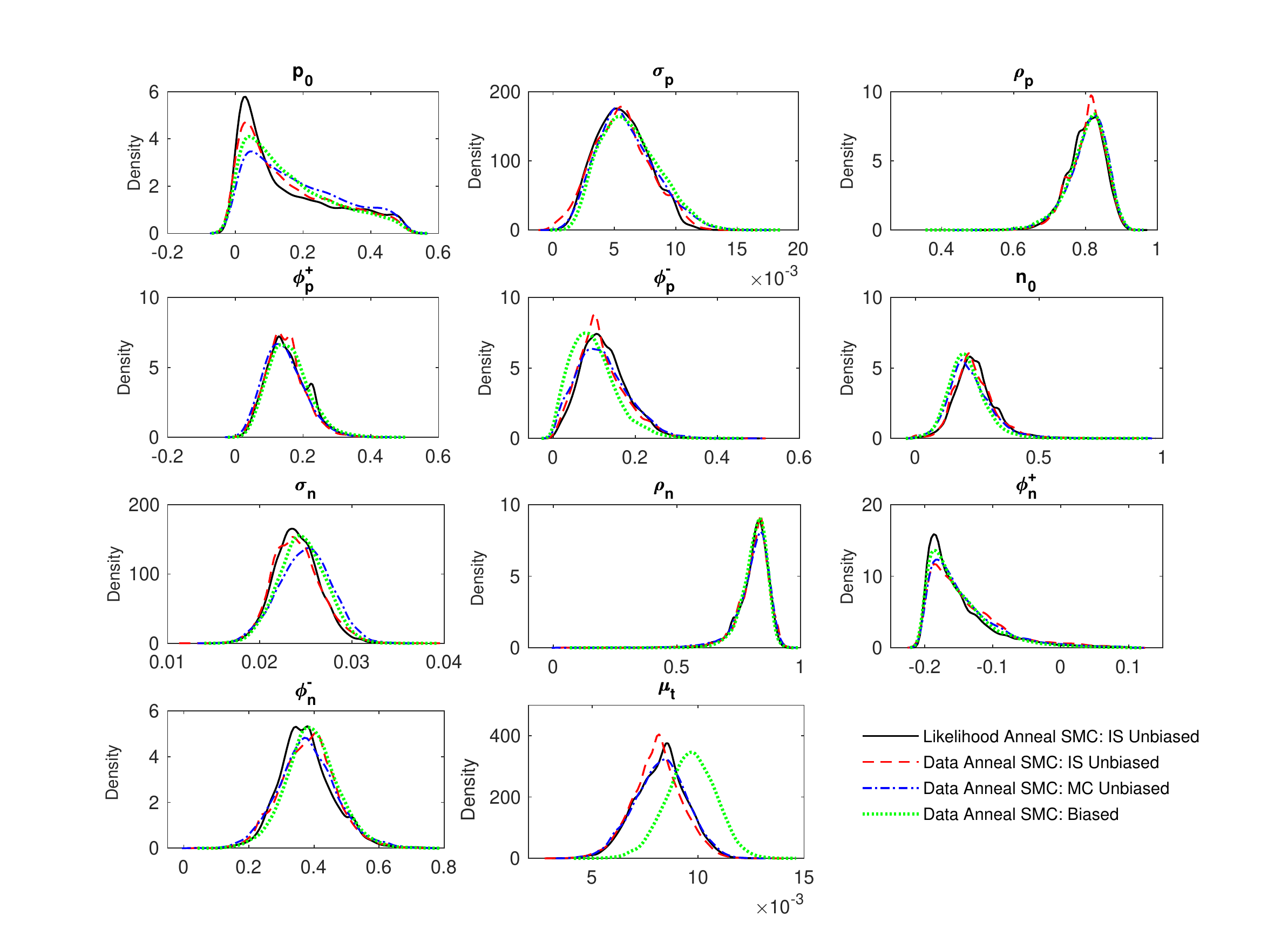}
	\captionsetup{font=small,skip=0pt}
	\caption{\small Marginal posterior distributions of the parameters of the BEGE model by using SMC.}
	\label{figure1}
\end{figure}
To test the efficiency of these two estimation methods, we choose three sets of parameters and 1000 independent likelihood estimates for the two estimators with different Monte Carlo sample sizes $M$. As shown in Table 1, the Monte Carlo integration requires to draw at least 2,000 times from a density to guarantee the standard deviation is lower than 1, while the importance sampling only needs 1,000 samples to achieve a similar variance. Therefore, we adopt the importance sampling estimator when using SMC to sample from the posterior for the BEGE model below.
\newcolumntype{L}{>{\centering\arraybackslash}m{3.3cm}}
\begin{table}[!htp]
	\centering
	\caption{\small Comparison of unbiased Monte Carlo integration estimator and importance sampling estimator} 
	\begin{tabular}{cLLL}
		\hline
		\hline
		& Number of Monte Carlo Iterations & \multicolumn{2}{c}{Standard Deviation}  \\
		\hline
		\diagbox{Parameters}{Methods}& & Monte Carlo Integration & Importance Sampling
		\\[0.2em]
		\hline
		\hline
		\multirow{4}{*}{Parameter Set 1} & {100} & 3.59 & 1.57\\
		&500& 1.61&1.04 \\
		&1000& 1.24& 0.91\\
		&2000& 0.94& 0.78\\
		&5000& 0.79& 0.70\\
		\hline
		\multirow{4}{*}{Parameter Set 2} & {100} & 3.13 & 1.51\\
		&500& 1.48& 0.87\\
		&1000& 1.13& 0.80\\
		&2000& 0.90& 0.64\\
		&5000& 0.74& 0.51\\
		\hline
		\multirow{4}{*}{Parameter Set 3} & {100} & 3.06 & 1.43\\
		&500& 1.33& 0.92\\
		&1000& 1.00& 0.79\\
		&2000& 0.76& 0.67\\
		&5000& 0.64& 0.49\\
		\hline
	\end{tabular}	
	\caption*{\footnotesize Parameter Set 1: $p_0=0.201$, $\sigma_p=0.008$, $\rho_p=0.800$, $\phi_p^+=0.141$, $\phi_p^-=0.214$, $n_0=0.241$, $\sigma_n=0.022$, $\rho_n=0.850$, $\phi_n^+=-0.167$, $\phi_n^-=0.215$, $\mu^\text{B}=0.009$;\\Parameter Set 2: $p_0=0.348$, $\sigma_p=0.004$, $\rho_p=0.706$, $\phi_p^+=0.280$, $\phi_p^-=0.129$, $n_0=0.186$, $\sigma_n=0.021$, $\rho_n=0.830$, $\phi_n^+=-0.086$, $\phi_n^-=0.347$, $\mu^\text{B}=0.010$;\\Parameter Set 3: $p_0=0.221$, $\sigma_p=0.009$, $\rho_p=0.710$, $\phi_p^+=0.223$, $\phi_p^-=0.175$, $n_0=0.275$, $\sigma_n=0.022$, $\rho_n=0.827$, $\phi_n^+=-0.195$, $\phi_n^-=0.380$, $\mu^\text{B}=0.008$.}
\end{table}

Both the MH MCMC and SMC methods are used to sample the posterior distributions for the BEGE model with simulated data, and the results are shown in Figure 2. The red dashed lines are the parameters used to generate the simulated data and are the initial values for the MCMC.
It is evident that the estimates of the posterior distribution produced by MH MCMC are less smooth than SMC and the values used to generate simulated data are also contained by the posterior distributions. This simulation study can be viewed as a validation of the effectiveness of the proposed SMC methods.

\begin{figure}[!h]
	\centering
	\includegraphics[width=1.05\linewidth, height=10cm]{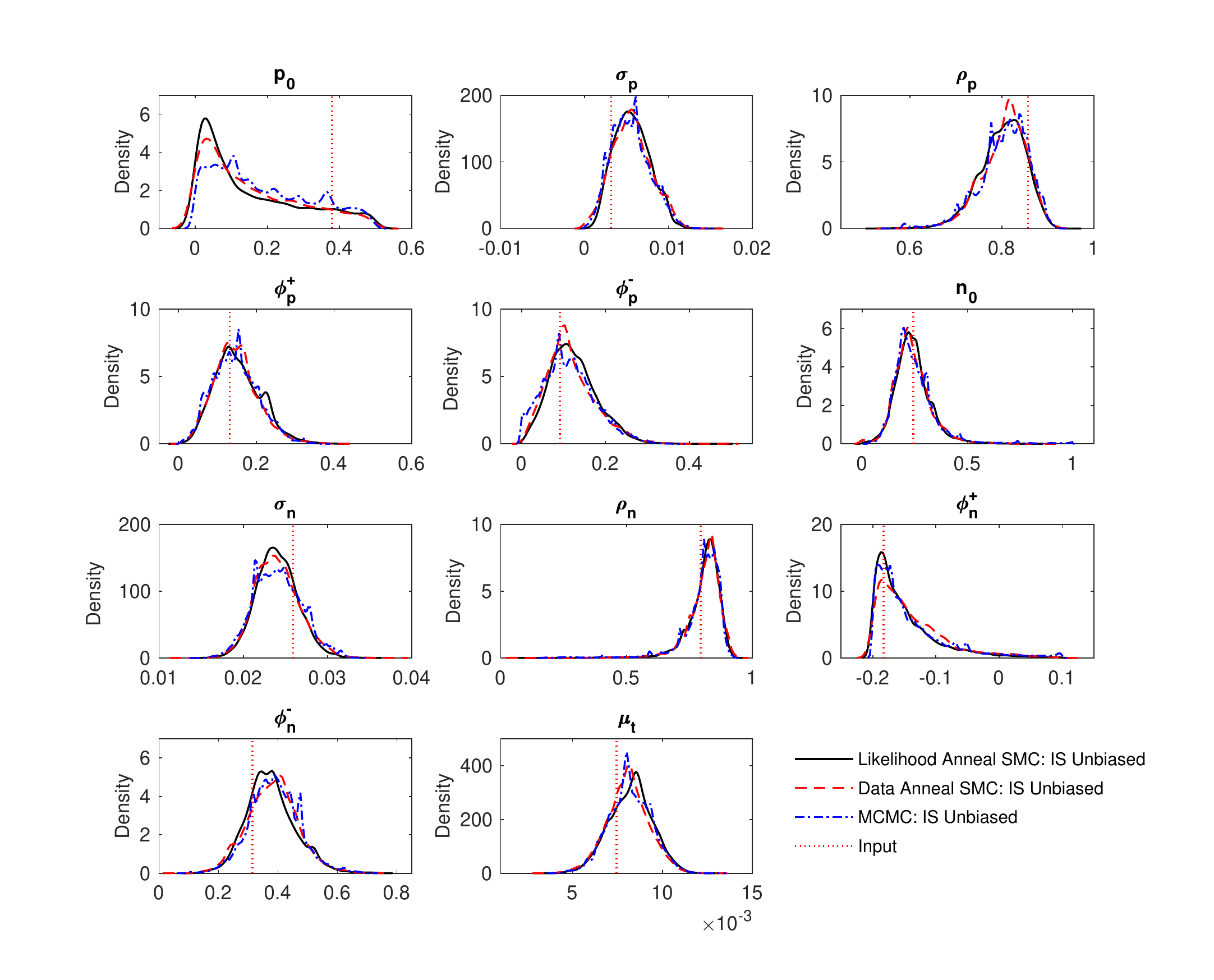}
	\captionsetup{font=small,skip=-5pt}
	\caption{\small Marginal posterior distributions of the parameters of the BEGE model by using SMC and MCMC methods for simulated data.}
	\label{figure2}
\end{figure}

\subsection{Comparison of MCMC and SMC}
In this section, the real data are used to compare the performance of the MH MCMC and SMC methods.
The likelihood function for the BEGE model is computationally expensive. We significantly accelerate the likelihood calculations for the proposed parameters of all particles in one iteration of the MCMC step of the SMC algorithm by taking advantage of parallel computing.  More specifically, we allocate a set of these proposed parameters to each core on a multi-core machine. Using the importance sampling estimator of the likelihood and 12 cores, the computation time for the SMC algorithm with data annealing applied to the BEGE model is 22.4 hours which runs 6.7 times faster than a completely serial SMC algorithm. Furthermore, the likelihood annealing SMC applied to the BEGE model runs 6.4 times faster than a completely serial SMC algorithm that takes 103.5 hours. Although this empirical application has focused on monthly log returns, the proposed approach is also feasible for larger datasets such as daily returns. However, in the context of daily returns and unbiased BEGE likelihood, the estimation with SMC can be computationally intensive, and the computation time may vary with different datasets.

Figure 3 shows the posterior distributions for the BEGE model explored by using MH MCMC and SMC methods. The MH MCMC requires tuning of the proposal distribution. To make implementing the MH MCMC easier, the final population of SMC particles are used to form the proposal distribution. More specifically, we use a multivariate normal random walk with a covariance estimated from the SMC particles.
However, even though we give MH MCMC an advantage by avoiding the tuning step, the posteriors obtained by MCMC still perform poorly compared with the ones from SMC methods in Figure 3. For instance, the posterior distribution for parameter $\sigma_p$ obtained by MCMC has an obvious unexpected bump, and the trace plot in Figure 4 shows that MCMC can get stuck and the Markov chain explores the parameter space slowly. The runtime of the MH MCMC algorithm with 1 million iterations is more than 2.2 times slower than the data annealing SMC that exploits parallel computing.
\begin{figure}[!h]
	\centering
	\includegraphics[width=1.05\linewidth, height=10cm]{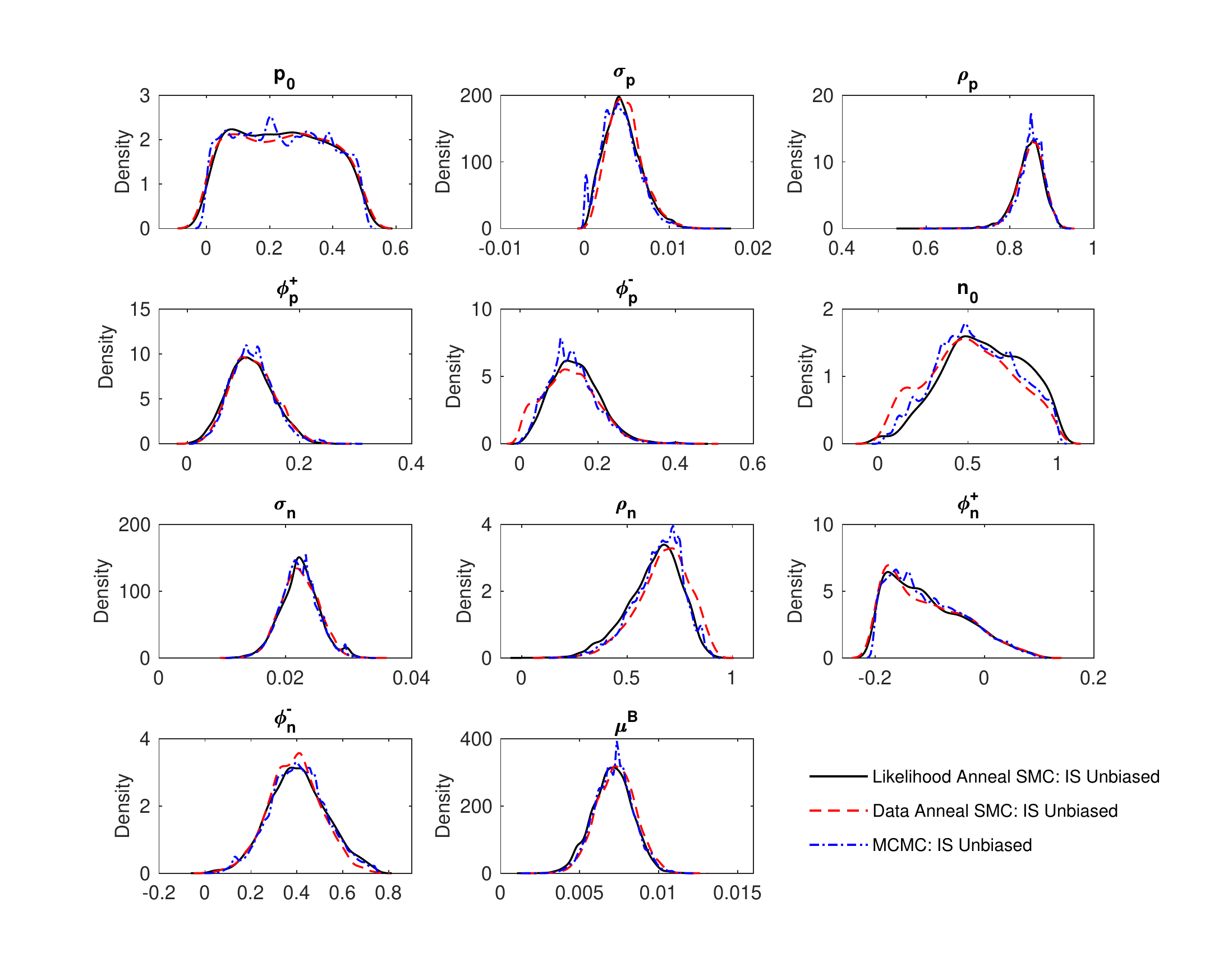}
	\captionsetup{font=small,skip=-10pt}
	\caption{\small Marginal posterior distributions of the parameters of the BEGE model by using SMC and MCMC methods for real data.}
	\label{figure3}
\end{figure}
\begin{figure}[!h]
	\hspace*{-1.3in}
	\includegraphics[width=1.4\linewidth, height=9cm]{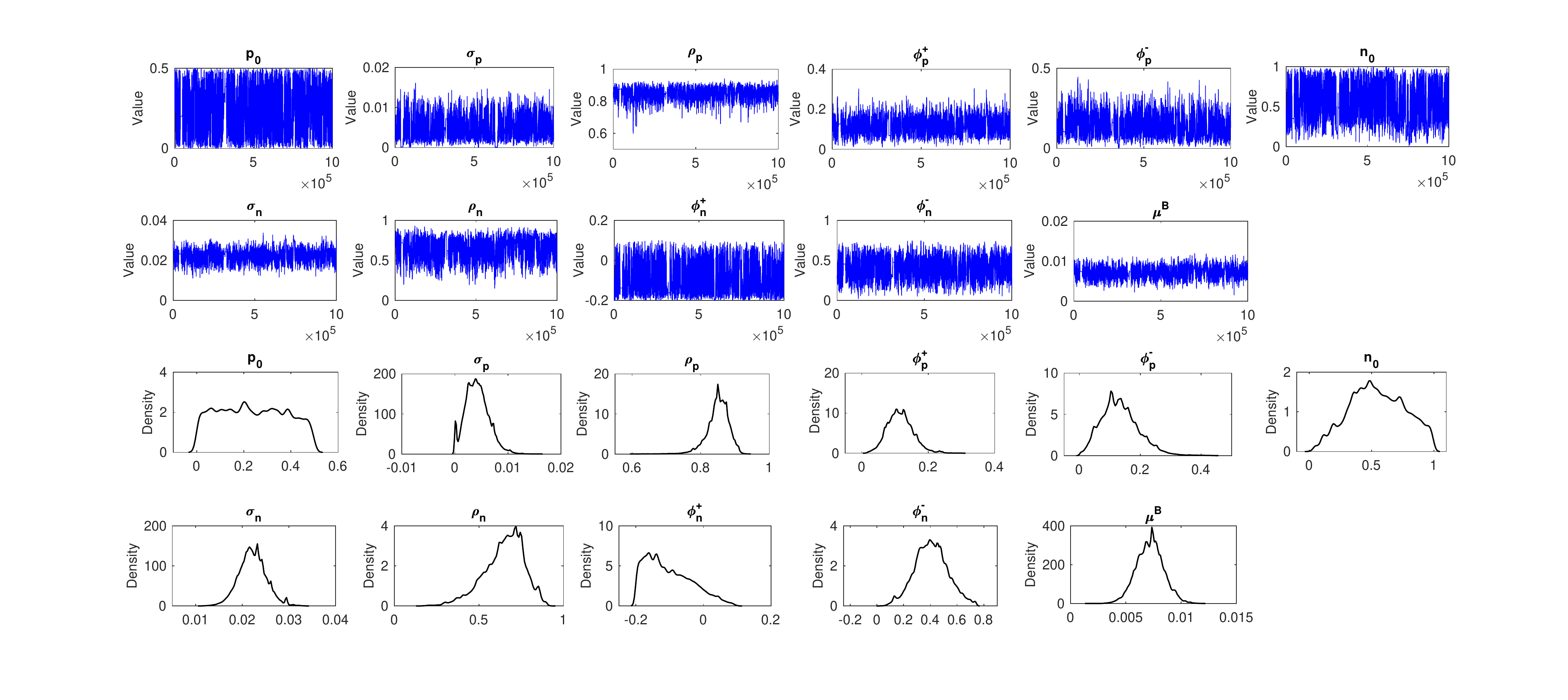}
	\captionsetup{font=small,skip=-15pt}
	\caption{\small Trace plots and marginal posterior distributions of the parameters of the BEGE model by using MCMC method for real data.}
	\label{figure4}
\end{figure}

The univariate posterior distributions obtained by our SMC algorithms for the parameters of the GARCH$(1,1)$ and GJR-GARCH$(1,1)$ are shown in Figures 5 and 6, respectively. As can be seen in the figures, each parameters' posterior distribution derived by data annealing SMC is very close to that obtained under likelihood annealing SMC. Except for the parameters of the fixed mean value $\bf{\mu}=(\mu^\mathrm{G}, \mu^{\mathrm{GJR}})$ under GARCH$(1,1)$ and GJR-GARCH$(1,1)$ model, all other posterior distributions demonstrate that they are not symmetric. Particularly, for the BEGE case in Figure 3, the posterior distributions are far from normally distributed, which contradicts the assumption of normally distributed parameter estimates under standard asymptotic inference.

\begin{figure}[!h]
	\centering
	\includegraphics[width=1.1\linewidth,height=6.5cm]{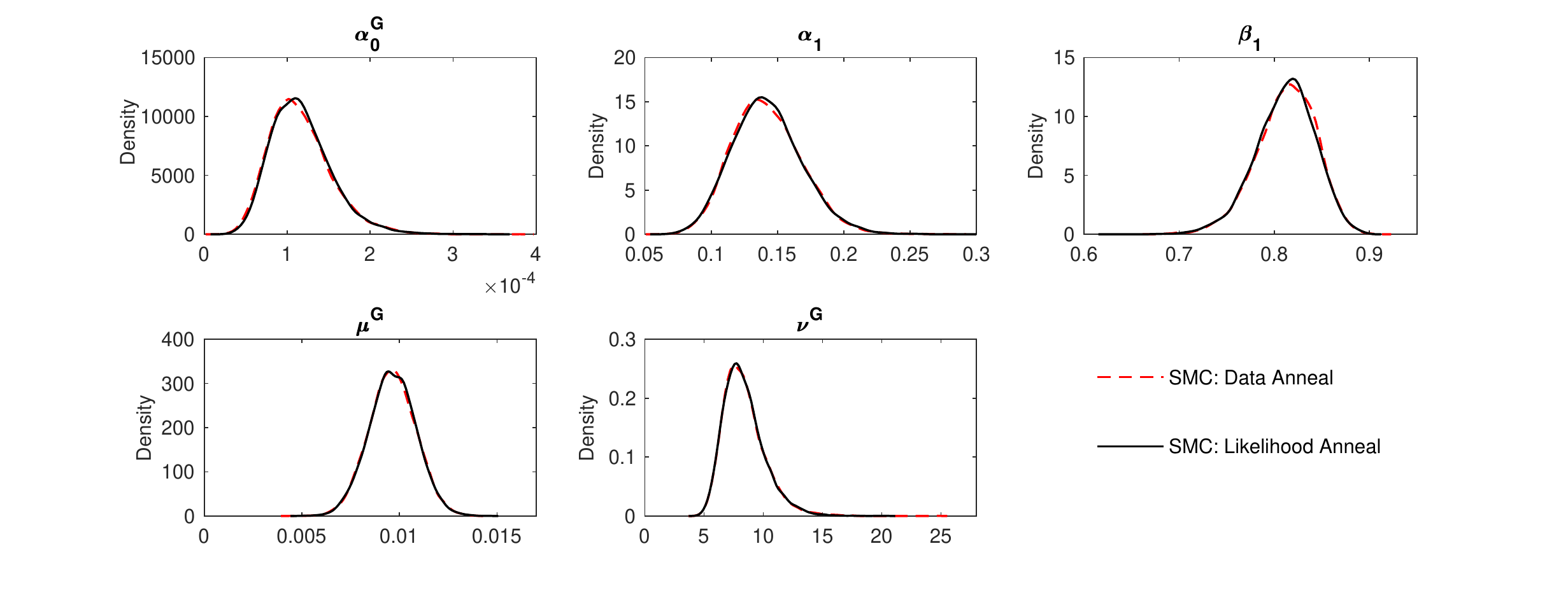}
	\captionsetup{font=small,skip=-5pt}
	\caption{\small Marginal posterior distributions of the parameters of the GARCH(1,1) model by using SMC.}
	\label{figure5}
\end{figure}

\begin{figure}[!h]
	\centering
	\includegraphics[width=1.1\linewidth, height=6.5cm]{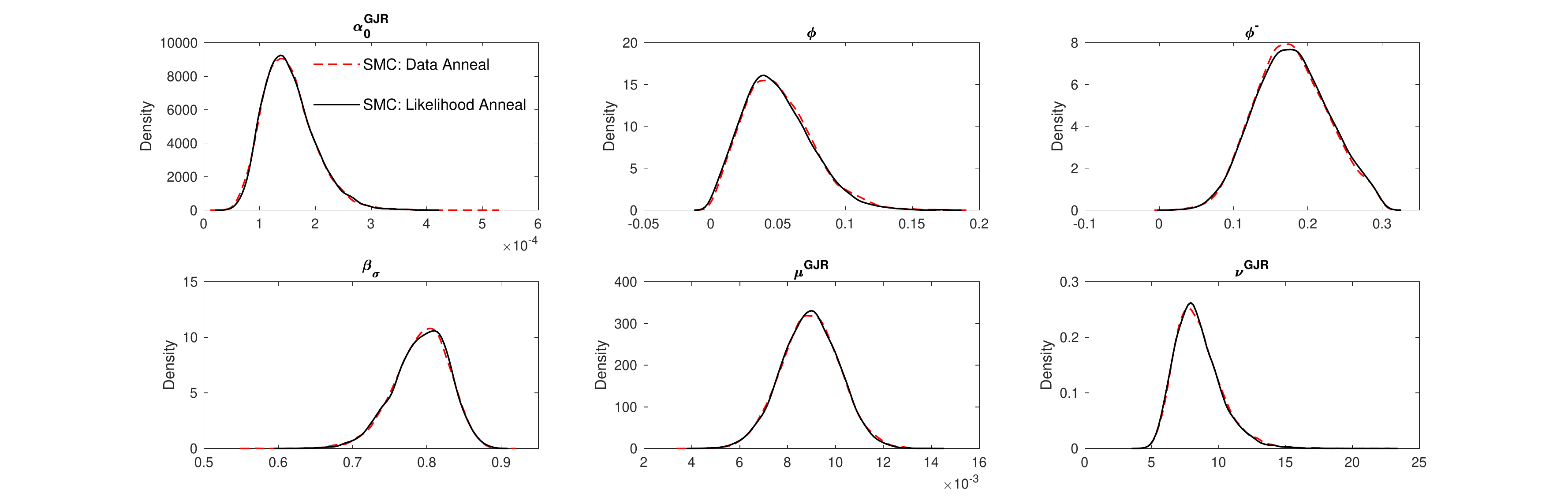}
	\captionsetup{font=small,skip=-5pt}
	\caption{\small Marginal posterior distributions of the parameters of the GJR-GARCH(1,1) model by using SMC.}
	\label{figure6}
\end{figure}

\subsection{Posterior Predictive Distribution}
\subsubsection{Simulation Study}
In this section we perform a simulation study which shows the relative merits of the Bayesian approach over the classical approach for forecasting the conditional variance. Firstly, we simulate the conditional variance of stock returns $\sigma_{1:1099}^2$ from the GARCH$(1,1)$ model, the GJR-GARCH$(1,1)$ model and the BEGE model, respectively. The parameters used to generate the three simulated data sets are randomly selected from the final population of SMC particles when modelling real time series data with the three GARCH-type models. The initial values of conditional variance at time $t=1$ of the three GARCH-type models are $(\sigma_1^2)_\mathrm{GARCH}=0.0023$, $(\sigma_1^2)_\mathrm{GJR-GARCH}=0.0022$ and $(\sigma_1^2)_\mathrm{BEGE}=\sigma_p^2p_1+\sigma_n^2n_1=0.0013$. The total number of simulated data in our study is 1099.  We consider the performance of one-step ahead prediction based on simulated conditional variances $t=201$ to $t=1099$. Besides using the true model to fit the simulated data, we also perform a study of model misspecification by using misspecified models to fit the simulated data and make predictions.
\begin{figure}[!h]
	\hspace*{-1in}
	\centering
	\includegraphics[width=1.5\linewidth, height=9.5cm]{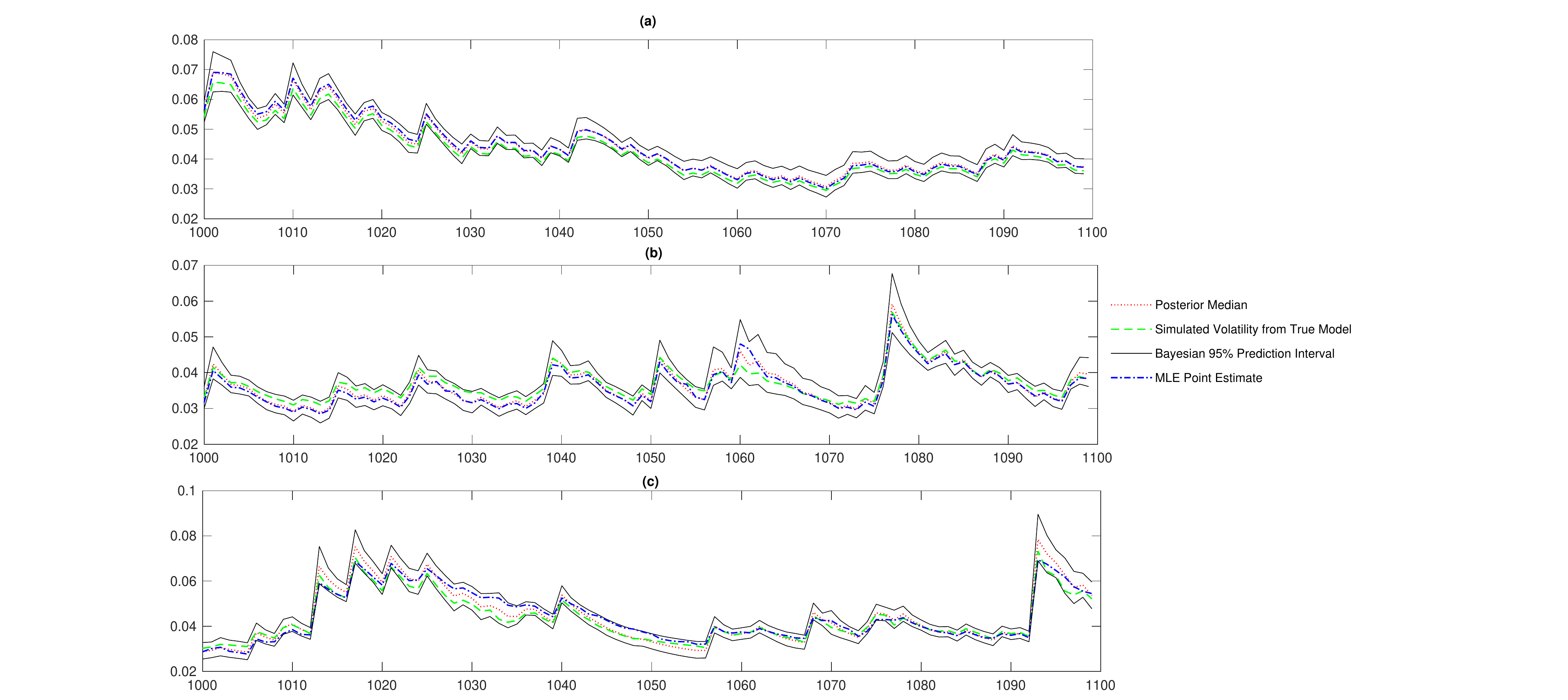}
	\captionsetup{font=small,skip=2pt}
	\caption{\small One-step ahead forecasts for conditional standard deviation of the GARCH$(1,1)$ model, the GJR-GARCH$(1,1)$ model and the BEGE model. (a) One-step ahead 95\% prediction interval of conditional standard deviation of GARCH$(1,1)$ model with simulated data generated from GARCH$(1,1)$ model; (b) One-step ahead 95\% prediction interval of conditional standard deviation of GJR-GARCH$(1,1)$ model with simulated data generated from GJR-GARCH$(1,1)$ model; (c) One-step ahead 95\% prediction interval of conditional standard deviation of BEGE model with simulated data generated from BEGE model.}
	\label{figure7}
\end{figure}

Here, we present the results for the one-step ahead forecasts of conditional variances of the three GARCH-type models obtained via data annealing SMC and MLE approaches. In order to provide a clearer visual interpretation, in Figure 7 we illustrate the results for forecasted conditional volatilities, which are the square roots of the forecasted conditional variances.
Figure 7 shows the 95\% prediction interval bounds for the last 100 predicted volatilities $\hat{\sigma}_{1000:1099}$ simulated from the three GARCH-type models  by using our Bayesian approach. The 95\% prediction interval is built simply by computing the 0.025 and 0.975 quantiles of the square roots of the posterior predictive sample of the conditional variances. However, the classical approach cannot produce such prediction intervals for the conditional volatility. It is obvious that these prediction intervals capture the simulated data successfully. We take posterior medians as point estimators for the Bayesian approach. After illustrating these posterior medians together with the classical point estimators of three GARCH-type models in Figure 7, we can see that the differences between the Bayesian and classical point estimators of predictions are small.

\newcolumntype{L}{>{\centering\arraybackslash}m{3.3cm}}
\begin{table}[!h]
	\centering
	\captionsetup{font=small,skip=1pt}
	\caption{\small Approximated $\mathrm{elpd}_{\mathrm{LFO}}$ for one-step ahead forecasts of simulated conditional variances from $t=201$ to $t=1099$} 
	\begin{tabular}{cLLL}
		\hline
		\hline
		\diagbox{Model}{Data} & {Simulated Data from GARCH$(1,1)$} & {Simulated Data from GJR-GARCH$(1,1)$}  & {Simulated Data from BEGE}\\
		\hline
		GARCH$(1,1)$ & {\textbf{6572.2}} & {3739.8} & {3069.1}\\
		\hline
		GJR-GARCH$(1,1)$ & {6352.1} & {\textbf{6505.7}}  & {5834.7}\\
		\hline
		BEGE & {6447.1} & {\textbf{6528.9}} & {\textbf{6188.1}}\\
		\hline
	\end{tabular}	
	
\end{table}

Each models' estimated $\mathrm{elpd}_{\mathrm{LFO}}$ for the one-step ahead forecasted conditional variance are given in Table 2. Besides checking the accuracy of Bayesian predictive distributions for conditional volatility by using the true model, we also use misspecified models to fit the simulated data and compare their performance with the true model. The model with the highest value of $\mathrm{elpd}_{\mathrm{LFO}}$ is the one that has the best predictive performance.
For the data simulated from the BEGE model, the log predictive density is much higher than the other two misspecified models that provide poor predictive performance. For the data simulated from the GARCH$(1,1)$ model the true model also provides the highest $\mathrm{elpd}_{\mathrm{LFO}}$.
It is worth noting that the BEGE model performs well for forecasting conditional variances from data simulated from the GJR-GARCH$(1,1)$ model. 
As the BEGE model allows for the type of asymmetries embedded within the GJR-GARCH$(1,1)$ model, it is to be expected that the BEGE model will perform well in this context. 
Therefore, even though the real volatility of the real stock returns is unobservable, this simulation study provides some confidence that the distribution of the real volatility can be adequately captured by adopting the Bayesian approach.

\subsubsection{Empirical Results}
With the advantages of the fully Bayesian approach demonstrated in the simulation study, we implement the methods to make forecasts for real data.
Figure 8 shows the time series for real stock returns and one-step ahead forecasts for conditional variances obtained via the Bayesian approach. The Bayesian posterior median value of the conditional variance is set as the point estimate for the one-step ahead prediction. It is immediately apparent that the Bayesian approach can provide efficient predictions from three GARCH-type models, which successfully capture the features of the data such as the evident growth of volatility following the negative shock of 1987 crash.
\begin{figure}[!h]
	\hspace*{-1in}
	\centering
	\includegraphics[width=1.3\linewidth, height=7.5cm]{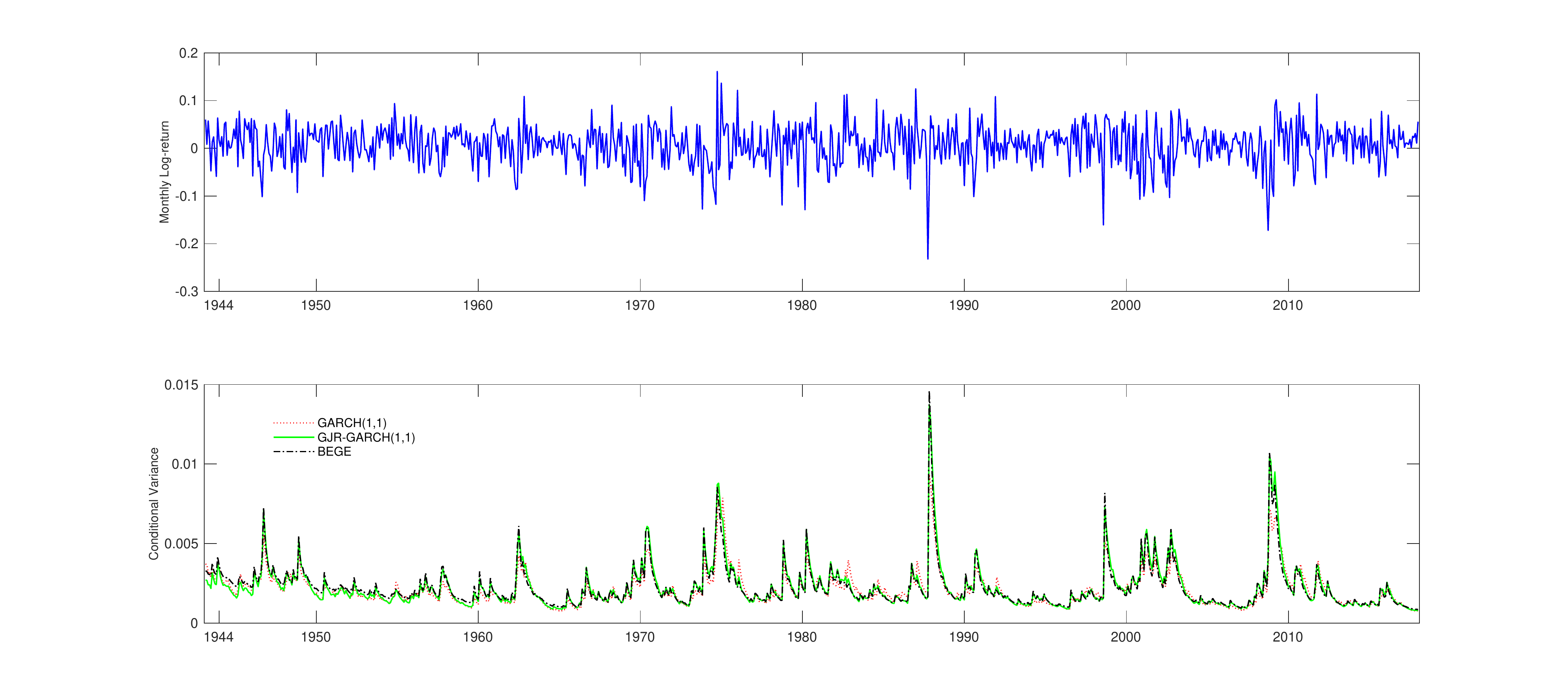}
	\captionsetup{font=small,skip=-5pt}
	\caption{\small One-step ahead Bayesian forecasts for conditional variance of GARCH$(1,1)$, GJR-GARCH$(1,1)$ and BEGE with real time series data by using the Bayesian approach. The prediction period is from March 1943 to January 2018. The top panel shows monthly log-return series, and the bottom panel shows the results of one-step ahead forecasted conditional variances from the three GARCH-type models.}
	\label{figure8}
\end{figure}

\newcolumntype{L}{>{\centering\arraybackslash}m{9cm}}
\begin{table}[!h]
	\hspace*{-0.5in}
	\centering
	\caption{\small Approximated $\mathrm{elpd}_{\mathrm{LFO}}$ for one-step ahead forecasts of real return from $t=201$ (March 1943) to $t=1099$ (January 2018)} 
	\begin{tabular}{cL}
		\hline
		\hline
		\diagbox{Model}{Data} & {Real Time Series Data} \\ 
		\hline
		GARCH$(1,1)$ & {1597.3}\\
		\hline
		GJR-GARCH$(1,1)$ & {1611.9}\\
		\hline
		BEGE & {\textbf{1613.6}} \\
		\hline
	\end{tabular}	
	
\end{table}

Table 3 provides each models' estimated $\mathrm{elpd}_{\mathrm{LFO}}$ for the one-step ahead predicted real return, where it is noted that the BEGE model has the highest predictive density indicating superior predictive performance. 
To examine this relative performance further, Figure 9 shows the log predictive density for each time step. For much of the sample, during periods of lower volatility, all three models exhibit, high, and similar values for the log predictive density. However, in most cases where volatility spikes and the log predictive density is lower, the GARCH$(1,1)$ and GJR-GARCH$(1,1)$ models consistently produce lower log predictive density than the BEGE model. This indicates that the superior performance of the BEGE model stems from periods of relatively high volatility.

\begin{figure}[!h]
	\hspace*{-1in}
	\centering
	\includegraphics[width=1.3\linewidth, height=6.5cm]{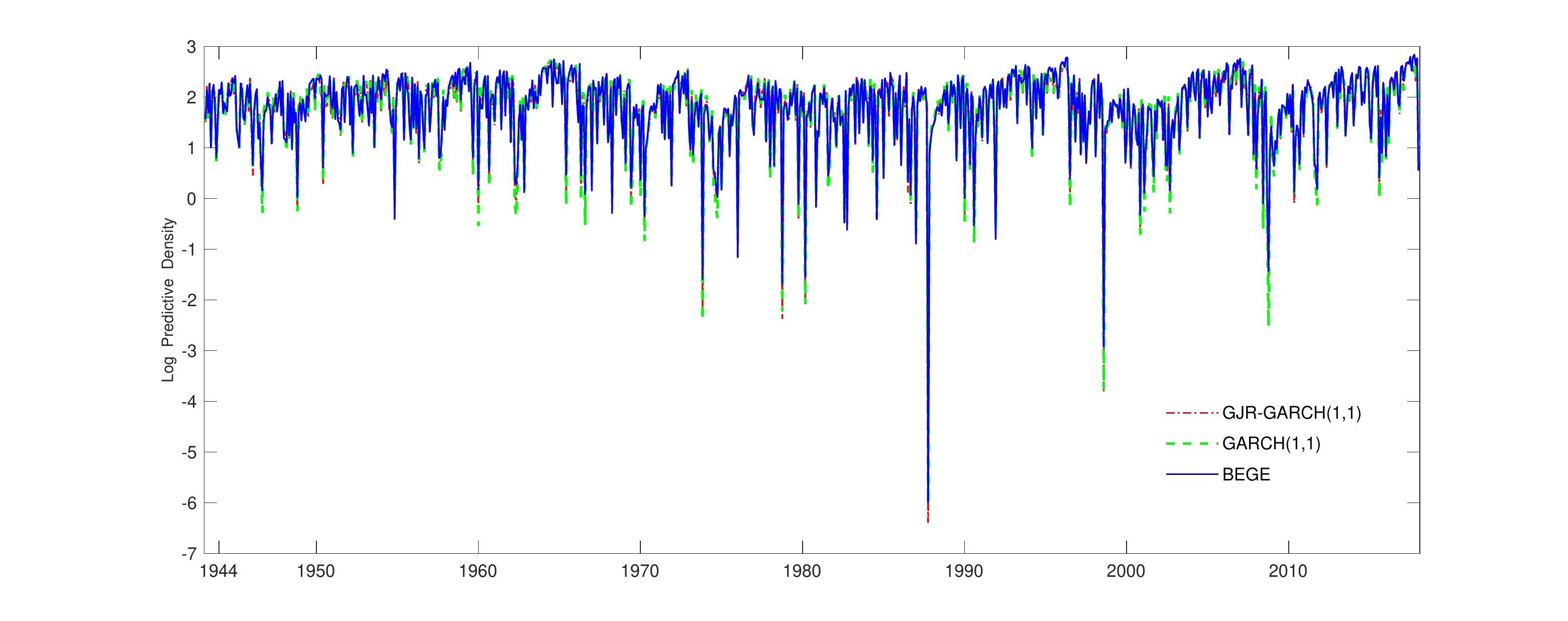}
	\captionsetup{font=small,skip=-5pt}
	\caption{\small Approximated log predictive density for one-step ahead forecasts of real return from $t=201$ (March 1943) to $t=1099$ (January 2018).}
	\label{figure9}
\end{figure}

\begin{figure}[h!]
	
	\hspace*{-1in}
	\centering
	\includegraphics[width=1.3\linewidth, height=8cm]{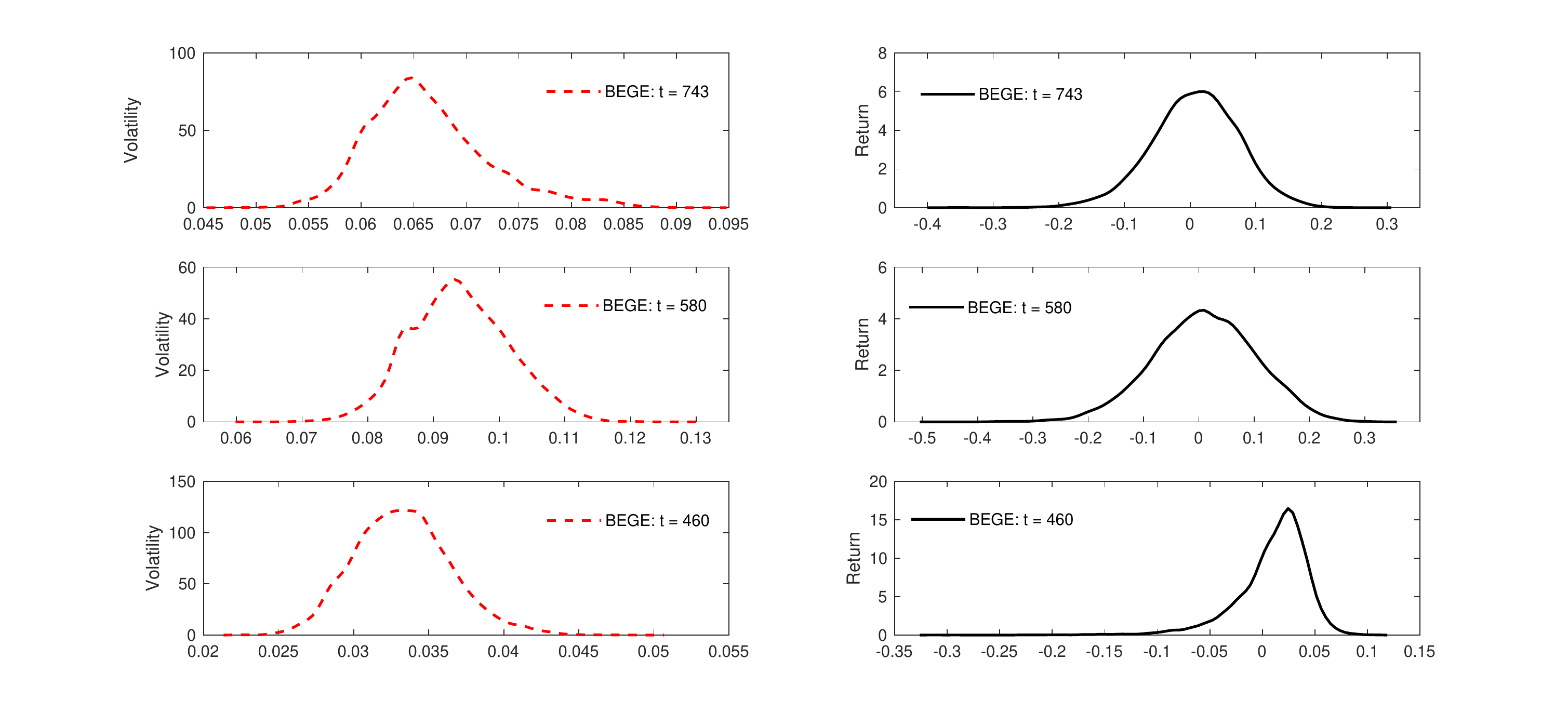}
	\captionsetup{font=small,skip=-5pt}
	\caption{\label{figure10} \small Predictive distribution for real data. The first column of panels report the distribution of one-step ahead predicted conditional volatility, and the corresponding predictive distributions of returns are shown on the right.} 
\end{figure}

Another promising aspect of the Bayesian approach is reflected in the provision of the predictive distribution of conditional volatility, which quantifies the uncertainty in the one-step ahead predicted volatility. However, this predictive distribution cannot be analytically derived.
In Figure 10, the first column shows some distributions of the predicted volatilities from the BEGE model, and the predictive distributions of returns are shown on the right.
From the predictive distribution, more information relating to the one-step forward unobserved volatility can be extracted. For instance, the probability of the conditional volatility at the next time point being within any particular range can be computed. The first plot shows a distribution of the predicted conditional volatility of the BEGE model at time $t=743$ conditional on real data $t=1$ to $t=742$. From this distribution, we can see that the conditional volatilities with values around 0.065 have a high probability of being observed. In contrast,
the last two plots on the left show that there is a lot of uncertainty about the relatively high and low volatility at time $t=580$ and $t=460$, respectively. For the corresponding distribution of predicted returns on the right, it is evident to see that the return at time $t=460$ is a relatively skewed distribution.

\subsection{Model Choice}
\subsubsection{Simulation Study}
In this section we perform a simulation study to assess the performance of the SMC estimator of the Bayesian model evidence for model selection for the GARCH-type models. We first simulate five sets of data from each competing model of interest, producing 15 datasets in total. The associated parameter values used to generate simulated data are randomly selected from the SMC posterior samples of the three GARCH-type models in Section 6.1. After obtaining simulated data, we use SMC to estimate the log evidence for each candidate model given different sets of simulated data.

\begin{table}[!h]
	\centering
	\captionsetup{font=small,skip=1pt}
	\caption{\small Estimation of Log Evidence - Simulated Data} 
	\begin{tabular}{cccccc}
		\hline
		\hline
		\\[-0.8em]
		& \multicolumn{5}{c}{Simulated Data from GARCH$(1,1)$}\\[0.2em]
		\hline
		\diagbox{Model}{Data} & 1 & 2 & 3 & 4 & 5 \\
		\hline
		GARCH$(1,1)$ & \textbf{1875.5} & \textbf{1898.6} & \textbf{1939.2} & \textbf{1947.4} & \textbf{1786.4} \\
		\hline
		GJR-GARCH$(1,1)$  & 1874.4 & 1898.4  & 1938.8 & 1945.2 & 1785.7 \\
		\hline
		BEGE & 1870.7 & 1895.2 & 1931.0 & 1940.3 & 1776.9 \\
		\hline
		\\[-0.8em]
		& \multicolumn{5}{c}{Simulated Data from GJR-GARCH$(1,1)$} \\[0.2em]
		\hline
		\diagbox{Model}{Data} & 1 & 2 & 3 & 4 & 5 \\
		\hline
		GARCH$(1,1)$ & 1889.0 & 2079.5 & 2002.3 & 2012.7 & 1929.6 \\
		\hline
		GJR-GARCH$(1,1)$  & \textbf{1895.1} & \textbf{2080.9}  & \textbf{2009.0} & \textbf{2018.2} & \textbf{1936.2} \\
		\hline
		BEGE & 1892.4 & 2080.3 & 2008.2 & 2015.1 & 1931.8 \\
		\hline
		\\[-0.8em]
		& \multicolumn{5}{c}{Simulated Data from BEGE} \\[0.2em]
		\hline
		\diagbox{Model}{Data} & 1 & 2 & 3 & 4 & 5 \\
		\hline
		GARCH$(1,1)$ & 1945.4 & 1862.4 & 1845.4 & 1892.2 & 1826.4 \\
		\hline
		GJR-GARCH$(1,1)$  & 1954.7 & 1869.0 & 1855.6 & 1899.1 & 1837.0 \\
		\hline
		BEGE & \textbf{1978.8} & \textbf{1881.4} & \textbf{1873.1} & \textbf{1915.7} & \textbf{1862.6} \\
		\hline
		
	\end{tabular}
\end{table}

The results of each models' estimation of log evidence given the simulated data are presented in Table 4. As the data generating process (DGP) is known, the performance of the evidence can be assessed by checking if the true model is correctly selected according to the estimated value of the log evidence. As can be seen from Table 4, for each simulated dataset the model with the highest estimated log evidence corresponds to the DGP. This demonstrates the effectiveness of the fully Bayesian approach for model choice and the SMC evidence estimator for the GARCH-type models and length of time series investigated here. It is worth noting that for simulated data from the GARCH$(1,1)$ model, the GJR-GARCH$(1,1)$ model has only a slightly smaller log evidence value compared to the DGP. However, when data is generated from the GJR-GARCH$(1,1)$ model, the log evidence of the GARCH$(1,1)$ model is much smaller than that of the DGP. When the GARCH$(1,1)$ model is true, there is a possibility that the GJR-GARCH$(1,1)$ model can mimic the behaviour of the GARCH$(1,1)$ model since the GARCH$(1,1)$ model is nested within the GJR-GARCH$(1,1)$ model.

\subsubsection{Empirical Results}
We also applied our methods to the real data.
Each models' estimated log evidence for the real time series data is provided in Table 5. There is strong evidence that the BEGE model is preferred over the other two GARCH-type models for this dataset. This is consistent with the result from \cite{bekaert2015bad} and reflects the fact that the BEGE model endowed with non-Gaussian innovations outperforms the traditional GARCH model and the asymmetric GJR-GARCH model, which are equipped with Student-t innovations, when applied to equity market returns. Relative to the traditional GARCH models, the BEGE framework offers a great deal more flexibility. Not only does the BEGE model allow for volatility components driving positive and negative returns, but the conditional shapes of the two distributions can also vary in much more complex ways that are impossible under more standard GARCH models. Different volatility processes for good and bad volatility and less restrictive distributional assumptions lead to a better explanation of observed returns.
\begin{table}[!h]
	\centering
	\captionsetup{font=small,skip=1pt}
	\caption{\small Estimation of Log Evidence - Real Data} 
	\begin{tabular}{cccccc}
		\hline
		\hline
		\\[-0.8em]
		\diagbox{Model}{Data} &  \multicolumn{5}{c}{Real Time Series Data} \\
		\hline
		& \multicolumn{2}{c}{Data Annealing SMC} &\multicolumn{3}{c}{Likelihood Annealing SMC} \\
		\hline
		GARCH$(1,1)$ & \multicolumn{2}{c}{1829.9} & \multicolumn{3}{c}{1830.1} \\
		\hline
		GJR-GARCH$(1,1)$  & \multicolumn{2}{c}{1839.4} & \multicolumn{3}{c}{1838.8}\\
		\hline
		BEGE & \multicolumn{2}{c}{\textbf{1845.4}} & \multicolumn{3}{c}{\textbf{1844.9}}\\
		\hline
	\end{tabular}
\end{table}

\begin{figure}[h!]
	\hspace*{-1in}
	\centering
	\includegraphics[width=1.3\linewidth, height=9cm]{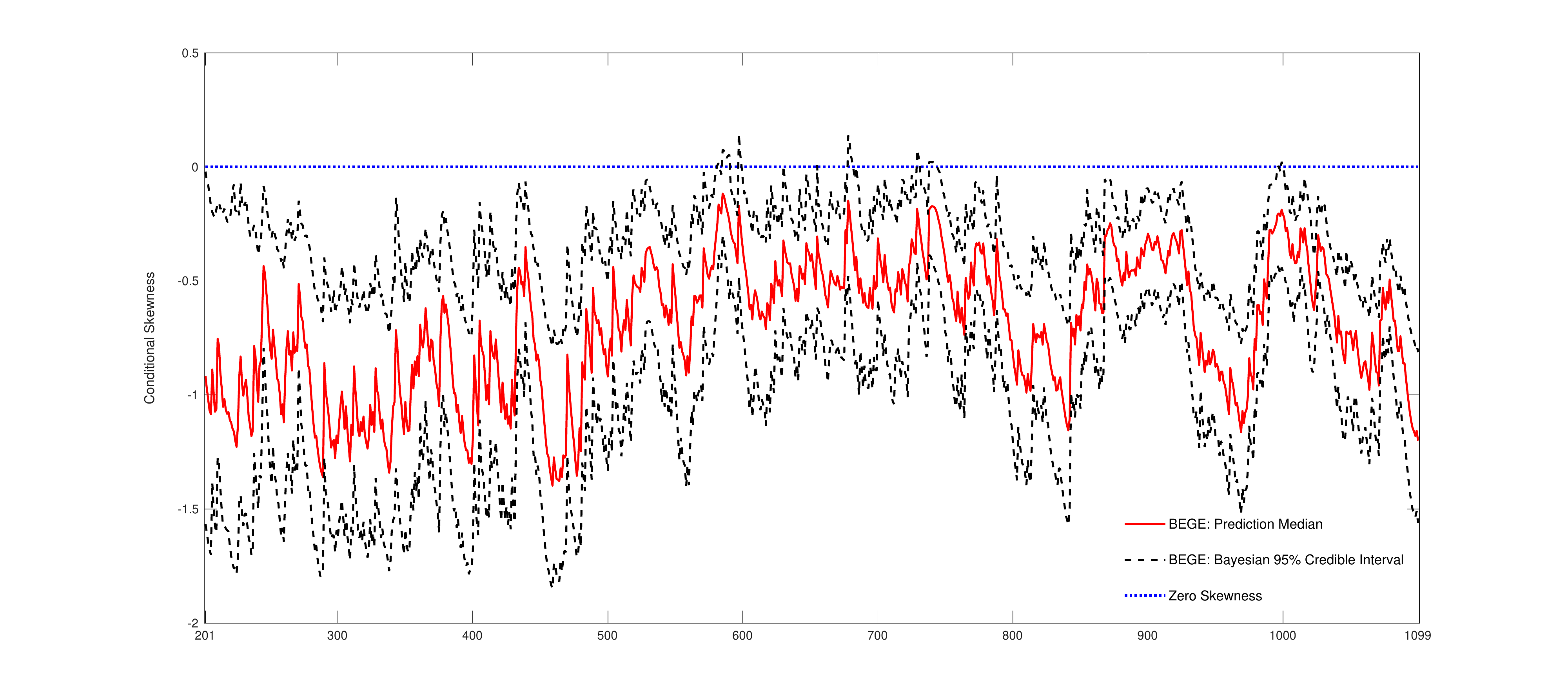}
	\captionsetup{font=small,skip=-5pt}
	\caption{\label{figure11} \small Predicted skewness for real data. The one-step ahead predicted skewness of the BEGE model and the constant zero skewness of the GARCH$(1,1)$ and GJR-GARCH$(1,1)$ model.}
\end{figure}

\begin{figure}[h!]
	\hspace*{-1in}
	\centering
	\includegraphics[width=1.3\linewidth, height=9cm]{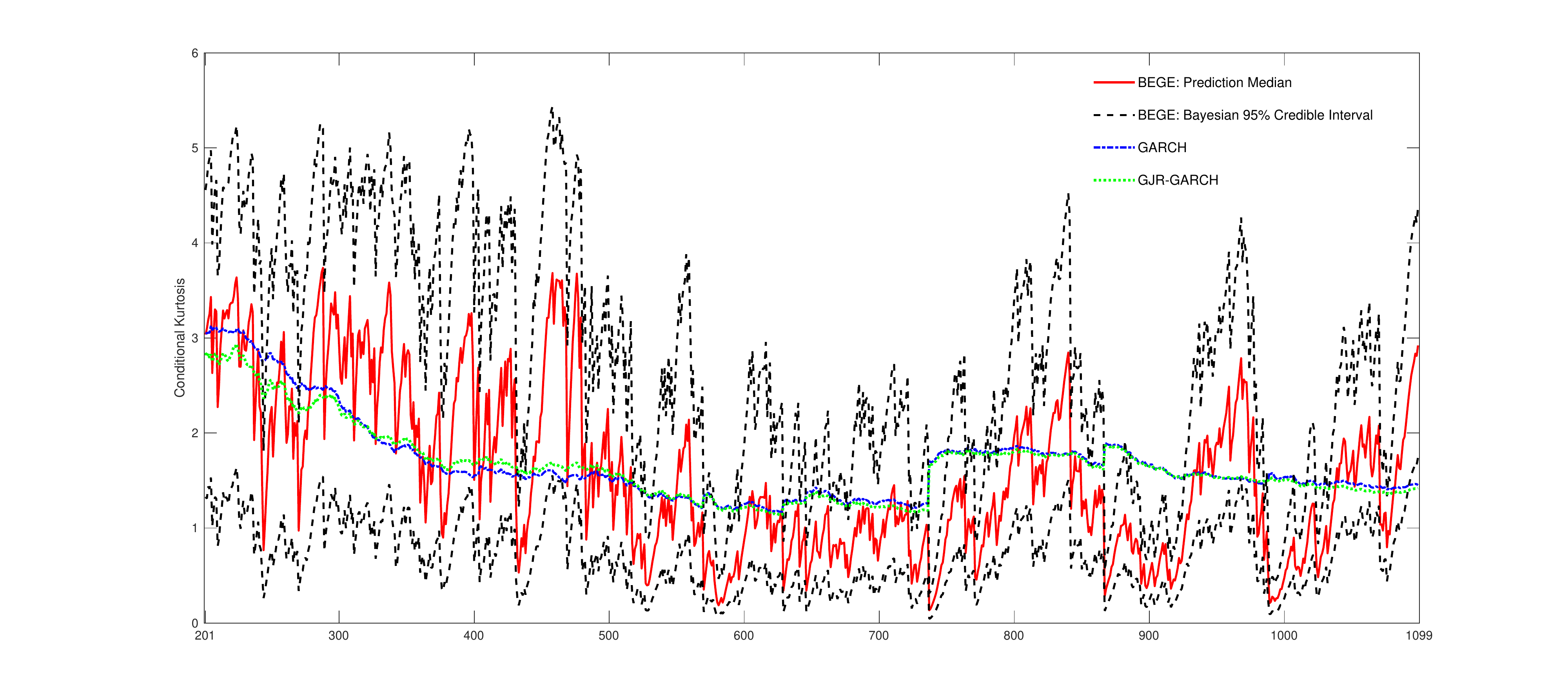}
	\captionsetup{font=small,skip=-5pt}
	\caption{\label{figure12} \small Predicted excess kurtosis for real data. The one-step ahead predicted excess kurtosis of BEGE, GARCH$(1,1)$ and GJR-GARCH$(1,1)$.}
\end{figure}

Moreover, the BEGE model can also produce the one-step ahead time-varying skewness and kurtosis. However, such dynamics for the higher conditional moments cannot be generated from the other candidate models with zero conditional skewness and constant conditional kurtosis. The time series of the median of the predicted conditional skewness of the BEGE model, together with the 95\% credible interval, is illustrated in Figure 11.
The time-varying skewness shows very different dynamics from the fixed zero skewness assumed in GARCH$(1,1)$ and GJR-GARCH$(1,1)$ for the prediction period from $t=201$ to $t=1099$. It demonstrates that the dynamic skewness is negative for this specific period, whereas the constant zero skewness is even above the upper bound of the credible interval most of the time. The time series of the predicted excess kurtosis for the same prediction period is shown in Figure 12. It is worth noting that the one-step ahead predicted excess kurtosis of GARCH$(1,1)$ and GJR-GARCH$(1,1)$, which is a function of the degrees of freedom of the assumed Student-t innovation, is not constant due to the changing of the parameters $\nu_t^{\mathrm{G}}$ and $\nu_t^{\mathrm{GJR}}$.
Therefore, more flexibility due to the two separate `good environment' component with positive skewness and the `bad environment' component with negative skewness, the richer time-varying higher-order moments and conditional non-Gaussianity captured by the BEGE model make this model outperforms the alternatives evidently when fitting the real data.

\section{Discussion}

In this article, a novel Bayesian framework for the analysis of GARCH-type models has been developed, that allows for parameter estimation, model selection and prediction. To this end, an efficient fully Bayesian approach has been developed using SMC. To obtain exact Bayesian inference for the BEGE model, an unbiased likelihood estimator was proposed to be used in place of the biased approximation adopted in \citet{bekaert2015bad}. This approach, in comparison to traditional estimation inference and forecasting techniques, offers a number of important advantages. Overall, in contrast to traditional methods, the Bayesian framework offers a direct method for quantifying the uncertainty surrounding both model parameter estimates and volatility forecasts. It is shown that SMC offers a number of advantages over standard Monte-Carlo methods. For such applications on long time series data, the SMC approach proposed here offers an efficient method for generating the full predictive distribution of volatility, computation of the evidence for model comparison and undertaking leave-one-out cross-validation. Although this study has focused on GARCH-type models, the benefits of the proposed approach would translate to other time series models for alternative economic or financial problems. We have demonstrated here that SMC data annealing can be an efficient approach for comparing the predictive performance of time series models. An interesting avenue for future research would be to explore this idea more comprehensively across different time series models and to draw comparisons with the approximate approach of \citet{burkner2019approximate}.

\section*{Acknowledgements}

The authors would like to thank the reviewers and associate editor for helpful comments to improve the paper.
We would also like to thank Leah South for the involvement in vectorising the code for the BEGE model and the corresponding implementation of SMC, which provided a good basis for developing this work. The first author was supported by a scholarship from the Australian Research Council Centre of Excellence for Mathematical and Statistical Frontiers (ACEMS). CD was supported by an Australian Research Council Discovery Project (DP200102101). Computational resources and services used in this work were provided by the HPC and Research Support Group, Queensland University of Technology, Brisbane, Australia.

\section*{References}
\renewcommand\refname{}
\bibliographystyle{apalike}
\bibliography{ref_test}

\section*{Appendix: Importance sampling estimator of likelihood of the BEGE model}
As described in Section 2.4.1, we need an importance distribution that will produce a low variance estimator of the likelihood with a small Monte Carlo sample size. This appendix shows that the likelihood requires to be estimated only when both $p_t > 1$ and $n_t >1$, and in other cases an exact likelihood can be computed. For simplicity, we only consider here the likelihood for a single observation. Four cases are considered to compute the likelihood $f_{\mathrm{BEGE}}(u)$ with the following expression
\begin{equation*}
\begin{aligned}
&f_{\mathrm{BEGE}}(u)=\int_{\omega_{p}}f_{\omega_{n}}(\omega_{p} - u)f(\omega_{p}) \mathrm{d}\omega_{p}, \mathrm{where}\\
&f_{\omega_{p}}(\omega_{p}) = \frac{1}{\Gamma(p) \sigma_{p}^{p}} \left(\omega_{p} - \overline{\omega}_{p}\right)^{p-1} \exp\left(-\frac{1}{\sigma_p}\left(\omega_{p} - \overline{\omega}_{p}\right)\right),  \mathrm{for} \enspace \omega_{p} > \overline{\omega}_{p}\\
&f_{\omega_{n}}(\omega_{p} - u) = \frac{1}{\Gamma(n) \sigma_{n}^{n}} \left(\omega_{p} - u - \overline{\omega}_{n}\right)^{n-1} \exp\left(-\frac{1}{\sigma_n}\left(\omega_{p} - u - \overline{\omega}_{n}\right)\right),  \mathrm{for} \enspace \omega_{p} - u > \overline{\omega}_{n}
\end{aligned}
\end{equation*}
where $p \ge 1$, $n \ge 1$, $\overline{\omega}_{p} = -p \sigma_p$, and $\overline{\omega}_{n} = -n \sigma_n$. The lower limit of integration in the expression for $f_{\mathrm{BEGE}}(u)$ is $\omega_{p}$, which must be greater than $\overline{\omega}_{p}$ and $\overline{\omega}_{n} + u$. Therefore, we define the lower limit of integration as $\delta = \max\left(\overline{\omega}_{p}, \overline{\omega}_{n} + u\right)$.

Firstly, define $\sigma = \frac{1}{\sigma_p}+\frac{1}{\sigma_n}$. When $p = n = 1$ an exact expression for $f_{\mathrm{BEGE}}(u)$ can be found by
\begin{equation*}
\begin{aligned}
f_{\mathrm{BEGE}}(u) &= \int_{\delta}^\infty \frac{1}{\sigma_p} \exp\left(-\frac{1}{\sigma_p}\left(\omega_{p} - \overline{\omega}_{p}\right)\right) \frac{1}{\sigma_n}\exp\left(-\frac{1}{\sigma_n}\left(\omega_{p} - (u + \overline{\omega}_{n})\right)\right) \mathrm{d}\omega_{p}\\
&= k_1\int_{\delta}^\infty\exp\left(-\sigma\omega_{p}\right)\mathrm{d}\omega_{p}\\
&= k_1\frac{1}{\sigma}\exp(-\sigma\delta), \mathrm{where} \enspace k_1 = \frac{1}{\sigma_p}\frac{1}{\sigma_n}\exp\left(\frac{\overline{\omega}_{p}}{\sigma_p}\right)\exp\left(\frac{u+\overline{\omega}_{n}}{\sigma_n}\right).
\end{aligned}
\end{equation*}
Then, the log-likelihood of the BEGE model in this case is
$\log{\left[f_{\mathrm{BEGE}}(u)\right]} = \log{k_1} - \log{\sigma} - \sigma\delta$.
When $p = 1$ and $n > 1$, an exact expression for the likelihood can be found as
\begin{equation*}
\begin{aligned}
&f_{\mathrm{BEGE}}(u) = \int_{\delta}^\infty \frac{1}{\sigma_p} \exp\left(-\frac{1}{\sigma_p}\left(\omega_{p} - \overline{\omega}_{p}\right)\right) \frac{1}{\Gamma(n) \sigma_{n}^{n}} \left(\omega_{p} - u - \overline{\omega}_{n}\right)^{n-1} \exp\left(-\frac{1}{\sigma_n}\left(\omega_{p} - u - \overline{\omega}_{n}\right)\right) \mathrm{d}\omega_{p},
\end{aligned}
\end{equation*}
define $k_2 = \frac{1}{\sigma_p } \frac{1}{\Gamma(n) \sigma_{n}^{n}}\exp\left(\frac{\overline{\omega}_{p}}{\sigma_p}\right)$, and $c_1 = k_2\exp\left(-\frac{1}{\sigma_p}\left(u + \overline{\omega}_{n}\right)\right)$, then the likelihood can be written as
\begin{equation*}
\begin{aligned}
&f_{\mathrm{BEGE}}(u) \\&= \int_{\delta}^\infty k_2 \exp\left(-\frac{1}{\sigma_n}\left(\omega_{p} - u - \overline{\omega}_{n}\right)\right) \exp\left(-\frac{\omega_{p}}{\sigma_p}+\frac{u+\overline{\omega}_{n}}{\sigma_p} - \frac{ u+\overline{\omega}_{n}}{\sigma_p }\right) \left(\omega_{p} - u - \overline{\omega}_{n}\right)^{n-1}\mathrm{d}\omega_{p}\\
&=\int_{\delta}^\infty c_1 \exp\left(-\sigma\left(\omega_{p} - u - \overline{\omega}_{n}\right)\right) \left(\omega_{p} - u - \overline{\omega}_{n}\right)^{n-1}\mathrm{d}\omega_{p}\\
&=c_1  \frac{\Gamma(n)}{\sigma^{n}} \int_{\delta}^\infty \frac{\sigma^{n}}{\Gamma(n)} \left(\omega_{p} - u - \overline{\omega}_{n}\right)^{n-1}\exp\left(-\sigma\left(\omega_{p} - u - \overline{\omega}_{n}\right)\right) \mathrm{d}\omega_{p}.
\end{aligned}
\end{equation*}
We set $x_{p} = \omega_{p} - u - \overline{\omega}_{n} = \omega_{p} - (u +\overline{\omega}_{n})$, which gives $\omega_{p} = x_{p} + u +\overline{\omega}_{n}$ and $\mathrm{d}\omega_{p} = \mathrm{d}x_{p}$. The likelihood then can be written as
\begin{equation*}
\begin{aligned}
f_{\mathrm{BEGE}}(u) &= c_1  \frac{\Gamma(n)}{\sigma^{n}} \int_{\delta - (u +\overline{\omega}_{n})}^{\infty} \frac{\sigma^{n}}{\Gamma(n)} x_{p}^{n-1} \exp\left(-\sigma x_{p}\right) \mathrm{d}x_{p}\\
&= \begin{cases}
c_1 \frac{\Gamma(n)}{\sigma^{n}}, & \text{if}\ \delta = u +\overline{\omega}_{n} \\
c_1 \frac{\Gamma(n)}{\sigma^{n}}\left[1-F(\overline{\omega}_{p} - (u +\overline{\omega}_{n}),n,\sigma)\right], & \text{if}\ \delta =\overline{\omega}_{p},
\end{cases}
\end{aligned}
\end{equation*}
where $F$ is the CDF of a gamma distribution with parameters $n$ and $\sigma$ over the interval of $\left(0,\overline{\omega}_{p} - (u +\overline{\omega}_{n})\right)$.
Similarly, in the case of $n=1$ and $p>1$ an exact expression for the likelihood can be found as
\begin{equation*}
\begin{aligned}
&f_{\mathrm{BEGE}}(u) = \int_{\delta}^\infty \frac{1}{\sigma_n} \exp\left(-\frac{1}{\sigma_n}\left(\omega_{p} - (u + \overline{\omega}_{n})\right)\right) \frac{1}{\Gamma(p) \sigma_{p}^{p}} \left(\omega_{p} - \overline{\omega}_{p}\right)^{p-1} \exp\left(-\frac{1}{\sigma_p}\left(\omega_{p} - \overline{\omega}_{p}\right)\right) \mathrm{d}\omega_{p},
\end{aligned}
\end{equation*}
define $k_3 = \frac{1}{\sigma_n} \frac{1}{\Gamma(p) \sigma_{p}^{p}}\exp\left(\frac{u+\overline{\omega}_{n}}{\sigma_n}\right)$ and $c_2 = k_3\exp\left(-\frac{\overline{\omega}_{p}}{\sigma_n}\right)$, then the likelihood can be written as
\begin{equation*}
\begin{aligned}
&f_{\mathrm{BEGE}}(u) \\&= \int_{\delta}^\infty k_3 \exp\left(-\frac{1}{\sigma_p}\left(\omega_{p} - \overline{\omega}_{p}\right)\right) \exp\left(-\frac{\omega_{p}}{\sigma_n}+\frac{\overline{\omega}_{p}}{\sigma_n} - \frac{\overline{\omega}_{p}}{\sigma_n }\right) \left(\omega_{p} - \overline{\omega}_{n}\right)^{p-1}\mathrm{d}\omega_{p}\\ &= \int_{\delta}^\infty c_2 \left(\omega_{p} - \overline{\omega}_{n}\right)^{p-1} \exp\left(-\sigma\left(\omega_{p} - \overline{\omega}_{p}\right)\right)\mathrm{d}\omega_{p}\\ &= c_2\frac{\Gamma(p)}{\sigma^{p}}\int_{\delta}^\infty \frac{\sigma^{p}}{\Gamma(p)} \left(\omega_{p} - \overline{\omega}_{n}\right)^{p-1} \exp\left(-\sigma\left(\omega_{p} - \overline{\omega}_{p}\right)\right) \mathrm{d}\omega_{p}.
\end{aligned}
\end{equation*}
We set $x_{p} = \omega_{p} - \overline{\omega}_{p} $, which gives $\omega_{p} = x_{p} +\overline{\omega}_{p}$ and $\mathrm{d}\omega_{p} = \mathrm{d}x_{p}$. The likelihood then can be written as
\begin{equation*}
\begin{aligned}
f_{\mathrm{BEGE}}(u) &= c_2  \frac{\Gamma(p)}{\sigma^{p}}\int_{\delta - \overline{\omega}_{p}}^{\infty} \frac{\sigma^{p}}{\Gamma(p)} x_{p}^{p-1} \exp\left(-\sigma x_{p}\right) \mathrm{d}x_{p}\\
&= \begin{cases}
c_2 \frac{\Gamma(p)}{\sigma^{p}}, & \text{if}\ \delta = \overline{\omega}_{p} \\
c_2 \frac{\Gamma(p)}{\sigma^{p}}\left[1-F(u +\overline{\omega}_{n}-\overline{\omega}_{p},p,\sigma)\right], & \text{if}\ \delta =u +\overline{\omega}_{n},
\end{cases}
\end{aligned}
\end{equation*}
where $F$ is the CDF of a gamma distribution with parameters $p$ and $\sigma$ over the interval of $\left(0,u +\overline{\omega}_{n}-\overline{\omega}_{p}\right)$.

In the case of $p>1$ and $n>1$, we use importance sampling method to estimate likelihood of the BEGE model. Here we attempt to find a parametric importance distribution $g(\omega_{p})$ that matched closely the target distribution $h(\omega_{p})$. Firstly, we determine the mode and curvature of $h(\omega_{p})$. The expression for $h(\omega_{p})$ is given by
\begin{equation*}
\begin{aligned}
h(\omega_{p}) = &\frac{1}{\Gamma(p) \sigma_{p}^{p}} \left(\omega_{p} - \overline{\omega}_{p}\right)^{p-1} \exp\left(-\frac{1}{\sigma_p}\left(\omega_{p} - \overline{\omega}_{p}\right)\right)\frac{1}{\Gamma(n) \sigma_{n}^{n}} \left(\omega_{p} - u - \overline{\omega}_{n}\right)^{n-1} \exp\left(-\frac{1}{\sigma_n}\left(\omega_{p} - u - \overline{\omega}_{n}\right)\right).
\end{aligned}
\end{equation*}
We set $m(\omega_{p}) = \log\left[h(\omega_{p})\right]$, which is given by
\begin{equation*}
\begin{aligned}
m(\omega_{p}) = &-\log{\Gamma(p)} - p\log{\sigma_p} + (p - 1)\log{(\omega_{p} - \overline{\omega}_{p})} -\frac{1}{\sigma_p}(\omega_{p} - \overline{\omega}_{p}) \\
&-\log{\Gamma(n)} - n\log{\sigma_n} + (n - 1)\log{(\omega_{p} - u - \overline{\omega}_{n})} - \frac{1}{\sigma_n}(\omega_{p} - u - \overline{\omega}_{n}).
\end{aligned}
\end{equation*}
The first and second derivatives of $m(\omega_{p})$ are given by
\begin{equation*}
\begin{aligned}
m'(\omega_{p}) = \frac{p-1}{\omega_{p}-\overline{\omega}_{p}} + \frac{n-1}{\omega_{p}-u-\overline{\omega}_{n}} - \sigma,\\
m''(\omega_{p}) = \frac{1-p}{(\omega_{p}-\overline{\omega}_{p})^2} + \frac{1-n}{(\omega_{p}-u-\overline{\omega}_{n})^2}.
\end{aligned}
\end{equation*}
The mode $\hat{\omega}_{p}$ of $m(\omega_{p})$ can be found by setting $m'(\omega_{p}) =0$, then we get
\begin{equation*}
\begin{aligned}
&(p-1)(\omega_{p}-u-\overline{\omega}_{n}) + (n-1)(\omega_{p}-\overline{\omega}_{p}) - \sigma(\omega_{p}-\overline{\omega}_{p})(\omega_{p}-u-\overline{\omega}_{n}) = 0,\\
&\sigma\omega_{p}^2 + \left[-\sigma(u+\overline{\omega}_{n}+\overline{\omega}_{p})+(2-n-p)\right]\omega_{p} + \left[\sigma\overline{\omega}_{p}(u+\overline{\omega}_{n})+(p-1)(u+\overline{\omega}_{n})+(n-1)\overline{\omega}_{p}\right]=0.
\end{aligned}
\end{equation*}
By setting $a = \sigma$, $b = -\sigma(u+\overline{\omega}_{n}+\overline{\omega}_{p})+(2-n-p)$, and $c = \sigma\overline{\omega}_{p}(u+\overline{\omega}_{n})+(p-1)(u+\overline{\omega}_{n})+(n-1)\overline{\omega}_{p}$, then we can write $a \omega_{p}^2 + b \omega_{p} + c =0$ and the mode $\hat{\omega}_{p} = \frac{- b +\sqrt{b^2 – 4ac}}{2a}$ can be easily derived. The variance of $m(\omega_{p})$ is estimated by using the second-derivative of $m(\omega_{p})$ around the mode $\hat{\omega}_{p}$, which is given by
\[\hat{V} = -\left[\frac{1-p}{(\hat{\omega}_{p}-\overline{\omega}_{p})^2} + \frac{1-n}{(\hat{\omega}_{p}-u-\overline{\omega}_{n})^2}\right]^{-1}.\]
We propose a gamma distribution, $X \sim \Gamma(i, j)$ for $x>0$, with mode $\hat{\omega}_{p}$ and variance $\hat{V}$ to be the importance distribution. The mode of this gamma distribution is $(i-1)j$ and the variance is $ij^2$. As $\omega_{p}>\delta$, we need to make the importance distribution greater than $\delta$ and a shift gamma distribution, $Y = X+\delta$ for $y > \delta$, is selected. As $x = y - \delta$, we can show that $\left|\frac{\mathrm{d}x}{\mathrm{d}y}\right| = 1$. Then an expression for the density of this importance distribution $g(y)$ is given by
\begin{equation*}
\begin{aligned}
g(y) = f_X(x(y)) \left|\frac{\mathrm{d}x}{\mathrm{d}y}\right| = \frac{1}{j^i\Gamma(i)}(y-\delta)^{i-1}\exp(- \frac{y-\delta}{j}),
\end{aligned}
\end{equation*}
where the mode of this distribution is $(i-1)j+\delta$ and the variance is $ij^2$. The value of parameter $i$ and $j$ can be derived by solving $(i-1)j+\delta  = \hat{\omega}_{p}$ and $ ij^2 =\hat{V}$, and the results are given by
\begin{equation*}
\begin{aligned}
i &= \frac{-b_i+\sqrt{b_i^2-4\hat{V}^2}}{2\hat{V}}, \mathrm{where}\ b_i = -2\hat{V} - (\hat{\omega}_{p}-\delta)^2\\
j &= \frac{\hat{\omega}_{p}-\delta}{i-1}.
\end{aligned}
\end{equation*}
After obtaining the importance distribution $g(\cdot)$, we can unbiasedly estimate $f_{\mathrm{BEGE}}(u)$ by following Equation 15. It is worth noting that this importance distribution can be used only when the mode $\hat{\omega}_{p}$ is greater than $\delta$, which is the lower limit of integration in the expression for $f_{\mathrm{BEGE}}(u)$. For this case, it can be proved that $\hat{\omega}_{p}$ is always greater than $\delta$, and the details of proof are shown below.
\begin{equation*}
\begin{aligned}
\text{(1)\ when}\ \delta &= \overline{\omega}_{p} > (u + \overline{\omega}_{n}),\\
(2a\delta+b)^2 &= (2\sigma\overline{\omega}_{p} +b)^2 \\&= b^2 + 4\sigma^2\overline{\omega}_{p}^2 + 4\sigma\overline{\omega}_{p}\left[-\sigma(u+\overline{\omega}_{n}+\overline{\omega}_{p})+(2-n-p)\right]\\&= b^2 + 4\sigma^2\overline{\omega}_{p}^2 - 4\sigma^2\overline{\omega}_{p}^2 - 4\sigma^2\overline{\omega}_{p}(u+\overline{\omega}_{n}) + 4\sigma\overline{\omega}_{p}(2-n-p)\\&= b^2 - \left[4\sigma^2(u+\overline{\omega}_{n})\overline{\omega}_{p} + 4\sigma(n-1) \overline{\omega}_{p} +4\sigma(p-1) \overline{\omega}_{p}\right]\\&= b^2 - 4ad, \mathrm{where}\ d = \sigma(u+\overline{\omega}_{n})\overline{\omega}_{p} + (n-1) \overline{\omega}_{p} +(p-1) \overline{\omega}_{p},\\
\because c &= \sigma(u+\overline{\omega}_{n})\overline{\omega}_{p}+(n-1)\overline{\omega}_{p}+(p-1)(u+\overline{\omega}_{n}), \text{and}\ \overline{\omega}_{p} > (u + \overline{\omega}_{n})\\
\therefore d&>c, 4ad>4ac, b^2 - 4ad < b^2 - 4ac, (2a\delta+b)^2 < b^2 - 4ac, 2a\delta+b <\sqrt{ b^2 - 4ac },\\ \delta &< \frac{- b +\sqrt{b^2 - 4ac}}{2a}\\ \delta&<\hat{\omega}_{p};\\
\text{(2)\ when}\ \delta &= (u + \overline{\omega}_{n}) > \overline{\omega}_{p},\\
(2a\delta+b)^2 &= (2\sigma(u + \overline{\omega}_{n}) +b)^2 \\ &= b^2+4\sigma^2(u + \overline{\omega}_{n})^2 + 4\sigma(u + \overline{\omega}_{n})\left[-\sigma(u+\overline{\omega}_{n}+\overline{\omega}_{p})+(2-n-p)\right]\\ &=b^2 + 4\sigma^2(u + \overline{\omega}_{n})^2 - 4\sigma^2(u + \overline{\omega}_{n})^2 - 4\sigma^2\overline{\omega}_{p}(u+\overline{\omega}_{n}) - 4\sigma(u + \overline{\omega}_{n})(n+p-2)\\&= b^2 - 4\sigma\left[\sigma(u+\overline{\omega}_{n})\overline{\omega}_{p} + (n-1) (u+\overline{\omega}_{n}) +(p-1) (u+\overline{\omega}_{n})\right]\\&= b^2 - 4ae, \mathrm{where}\ e=\sigma(u+\overline{\omega}_{n})\overline{\omega}_{p} + (n-1) (u+\overline{\omega}_{n}) +(p-1) (u+\overline{\omega}_{n}),\\
\because c &= \sigma(u+\overline{\omega}_{n})\overline{\omega}_{p}+(n-1)\overline{\omega}_{p}+(p-1)(u+\overline{\omega}_{n}), \text{and}\ (u + \overline{\omega}_{n}) > \overline{\omega}_{p}\\
\therefore e&>c, 4ae>4ac, b^2 - 4ae < b^2 - 4ac, (2a\delta+b)^2 < b^2 - 4ac, 2a\delta+b <\sqrt{ b^2 - 4ac },\\ \delta &< \frac{- b +\sqrt{b^2 - 4ac}}{2a}\\ \delta&<\hat{\omega}_{p}.
\end{aligned}
\end{equation*}

\end{document}